\documentclass{cernyrep}
\usepackage{graphicx}
\usepackage{amssymb}
\usepackage[T1]{fontenc}
\usepackage[bookmarks, colorlinks=true, linktoc=page, pdftex, linkcolor=black, citecolor=black, urlcolor=blue]{hyperref}
\usepackage{tikz}
\usepackage{tikz-3dplot}
\usetikzlibrary{shapes,arrows}
\usetikzlibrary{positioning}
\tikzstyle{block} = [rectangle, draw, fill=blue!20, 
    text width=5em, text centered, rounded corners, minimum height=4em]
\tikzstyle{line} = [draw, -latex']

\usepackage{fancyhdr}
\fancyhfoffset{4 mm}
\fancypagestyle{ARTTITLE}{%
\fancyhf{} 
\chead{\small Proceedings of the 2017 CERN--Accelerator--School course on \it Vacuum for Particle Accelerators, Glumsl\"ov, (Sweden)} 
\lfoot{Available online at \url{https://cas.web.cern.ch/previous-schools}}
\rfoot{\thepage\hspace*{3mm}}

}

\begin{document}
\title{Impedances and Instabilities}

\author{R.~Wanzenberg}

\institute{Deutsches Elektronen-Synchrotron DESY, D-22603 Hamburg, Germany}

\newcommand{\IN}{ {\sl I\kern-.1667em{N}} }
\newcommand{\0}{ \phantom{0} }
\newcommand{\VEC}[1]{\mbox{ \boldmath $\hspace*{-1mm} #1 \hspace*{-1mm} $ } }
\newcommand{\Abs}[1]{ \left| #1 \right|}
\newcommand{\FT} [1]{ \widetilde{#1}}
\newcommand{\rd}{{\rm d}}
\newcommand{\ri}{{\rm i}}
\newcommand{\re}{{\rm e}}


\begin{abstract}
The concepts of wake fields and impedance are introduced
to describe the electromagnetic interaction of a bunch of charged particles
with its environment in an particle accelerator.
The wake fields can  act back  on  the  beam  and
lead to instabilities, which may limit the achievable
current  per  bunch,  the  total  current,  or  even  both.
Some typical examples are used to illustrate the wake function and its basic properties.
Then the frequency-domain view of the wake field
or impedance is explained, and basic properties of the impedance
are derived. The impedance of a cavity mode is illustrated using
an equivalent circuit model.
The relation of wake field effects to important beam parameters
is treated in the rigid beam approximation. Several examples are
employed to illustrate the impact of the geometry and the material
properties of the vacuum chamber on the impedance.
Finally, a basic introduction to beam instabilities based on a
head tail model of the beam is given.
\end{abstract}


\maketitle

\thispagestyle{ARTTITLE}

\section{Introduction}
A beam in an accelerator interacts  with  its
vacuum chamber surroundings via electromagnetic fields.
In this lecture the concept of wake fields is introduced
to describe the electromagnetic interaction of a bunch of charged particles
with its environment.  The various components of the environment are the vacuum
chamber, cavities, bellows, dielectric-coated pipes, and other kinds of
obstacles the beam has to pass on its way through the accelerator.
The wake fields can  act  back  on  the  beam  and
lead to instabilities, which may limit the achievable
current  per  bunch, the  total  current,  or even  both.

This lecture builds upon a previous lecture on wake fields and
impedance given by T.~Weiland  \cite{Wei92} and M.~Dohlus and the author \cite{Doh2015}.
Furthermore, there are excellent textbooks available
\cite{Chao93, Zotter:1998wt, Ng:2006eu, Wolski:2014lba} which cover the subject matter of this lecture.

In the introduction several typical examples are presented  which demonstrate the interaction
of a beam with the surrounding vacuum chamber via wake fields.

Then, in Section 2, the concept of wake potential is formally introduced
and multipole expansions are studied for structures with cylindrical symmetry.
The Panofsky--Wenzel theorem, which links the longitudinal and transverse wake forces,
is explained.

Section 3 is devoted to the analysis of wake fields due to resonant modes
in a cavity.  The coupling of the beam to one mode of a cavity leads
to the concept of the loss parameter.

In Section 4, the impedance is introduced as the Fourier transform of the
wake potential. The properties of the wake functions (time-domain view)
are translated to  properties of the impedance (frequency-domain view).

\subsection{Basic concepts}
Consider a point charge $ q $ moving in free space at a velocity $ v $
close to the speed of light.
The electromagnetic field is Lorentz-contracted into a thin disk
perpendicular to the particle's direction of motion \cite{Jackson}, which we choose to be
the $z$-axis in a cylindrical coordinate system. The opening angle of
the field distribution is of the order of $ 1/\gamma $, where
$ \gamma = {( 1 - {(v/c)}^2 )}^{-1/2} $.
The field distribution is shown in Fig.~\ref{fig01}.
Even for an electron beam with an energy of $E = 10\:{\rm MeV}$, the opening
angle $\phi$ is no greater than $50\:{\rm mrad}$ or $2.89^{\circ}$:
$$ \phi = \frac{1}{\gamma} = \frac{0.511 \, {\rm MeV}}{E} = 2.89^{\circ} $$
($m_0 c^2 = 0.511 \, {\rm MeV}$ is the rest mass of the electron).
\setlength{\unitlength}{1cm}
\begin{figure}[thbp]
  \begin{center}
    \begin{picture}(5,4)(0,0)
      \put(0,0){\rotatebox{0}{\resizebox{0.3\textwidth}{!}{
        \includegraphics*{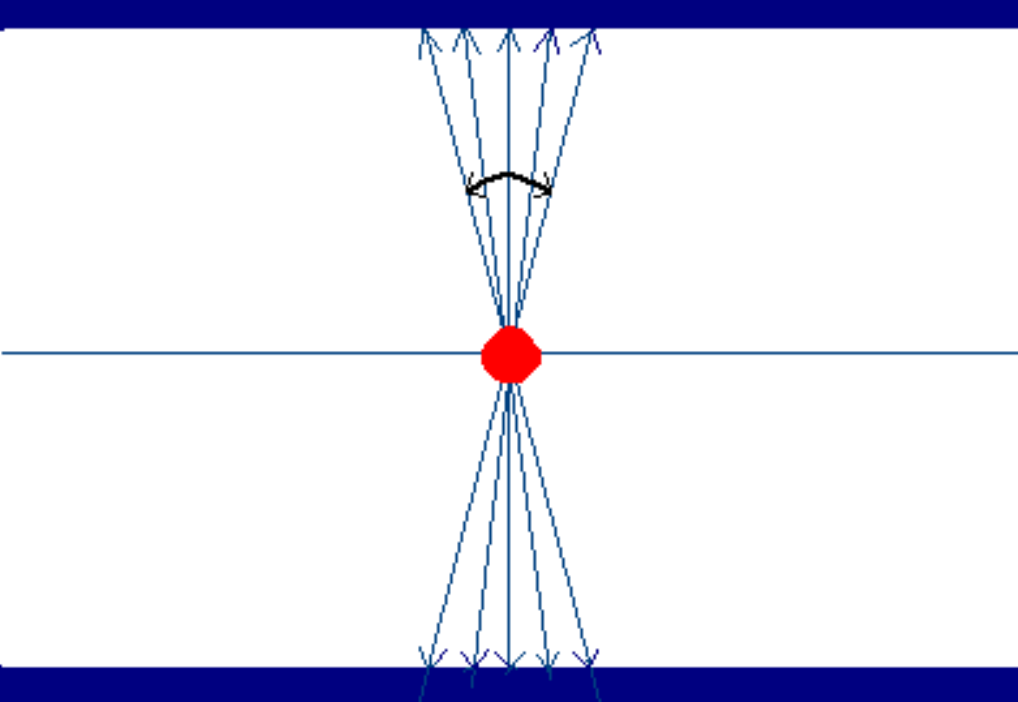}}}}
      \put(2.7,2.3){\makebox(0,0)[bl]{ $ \phi $}}
    \end{picture}
  \end{center}
\caption{
Electromagnetic field carried by a relativistic point charge $q$
 }
\label{fig01}
\end{figure}

In the ultra-relativistic limit $ v \rightarrow c $ (or
$ \gamma \rightarrow \infty $), the disk containing the field shrinks to
a $ \delta $-function distribution. The non-vanishing field
components are
$$
E_r = \frac{q}{2\pi \, \varepsilon_0 r} \,\delta(z-c\,t),
\qquad
H_{\varphi} = \frac{E_r}{Z_0}
\quad
\text{with } Z_0 = 377 \, \Omega.
$$

Since the electric field $ \VEC{E} $ points strictly radially outward from
the point charge, all field components are identically zero both ahead of and
behind the point charge, and hence there are no forces on a test
particle either preceding or following the charge~$ q $.

For $ v $ slightly less than $ c $, this is not strictly true. However, if we
look at some typical bunch charges and energies of high-energy accelerators and synchrotron light sources, as shown in Table~\ref{tab1-1}, we will notice that the space charge force
$ V_{\rm s} = e / (4 \, \pi \, \epsilon_0 \, d^2 \, \gamma^2) $
(where $ d $ is the rms distance between two electrons in the bunch)
scales with $1/\gamma^2$. It is then obvious that as a good approximation, any space charge effects
can be neglected for the  accelerators under consideration.
Nevertheless, space charge effects are important in heavy ion or
low-energy proton accelerators.
\begin{table}[h]
\def~{\hphantom{0}}
\begin{center}
\caption{Typical bunch charges and energies of high-energy accelerators and synchrotron light sources \cite{LHC_review,LEP_design,PETRAIII_TDR}} \label{tab1-1}
\begin{tabular}{lccc}
\hline
\hline
 Machine & {Charge (nC)}  & Energy (GeV) & $ \gamma = {( 1 - {(v/c)}^2 )}^{-1/2}$ \\
\hline
  LHC          & ~20          & 7000        &          ~~7\,500                     \\
  LEP          & 100         &   ~~60        &        195\,700                     \\
  PETRA III    & ~20          &    ~~~6        &         ~11\,700                     \\
\hline
\end{tabular}
\end{center}

\end{table}

In the next section we will restrict ourselves to the ultra-relativistic
case ($\gamma = \infty, \, v = c $), so space charge effects are
neglected.

\subsection{Some simple examples}
\begin{figure}[thbp]
   \begin{center}
    \rotatebox{270}{\resizebox{0.15\textwidth}{!}{%
          \includegraphics*[viewport=0 0 142 643,clip]{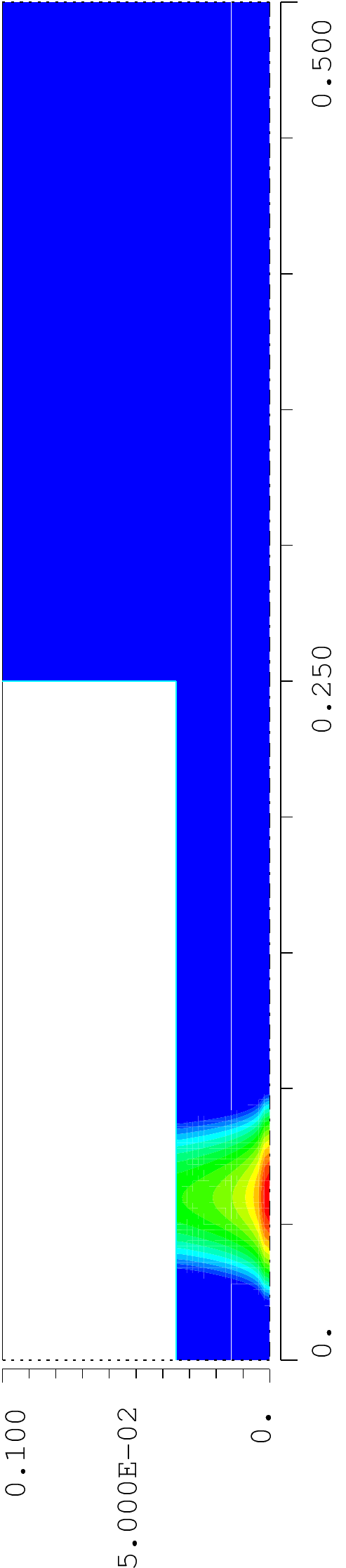}}}

    \rotatebox{270}{\resizebox{0.15\textwidth}{!}{%
          \includegraphics*[viewport=0 0 142 643,clip]{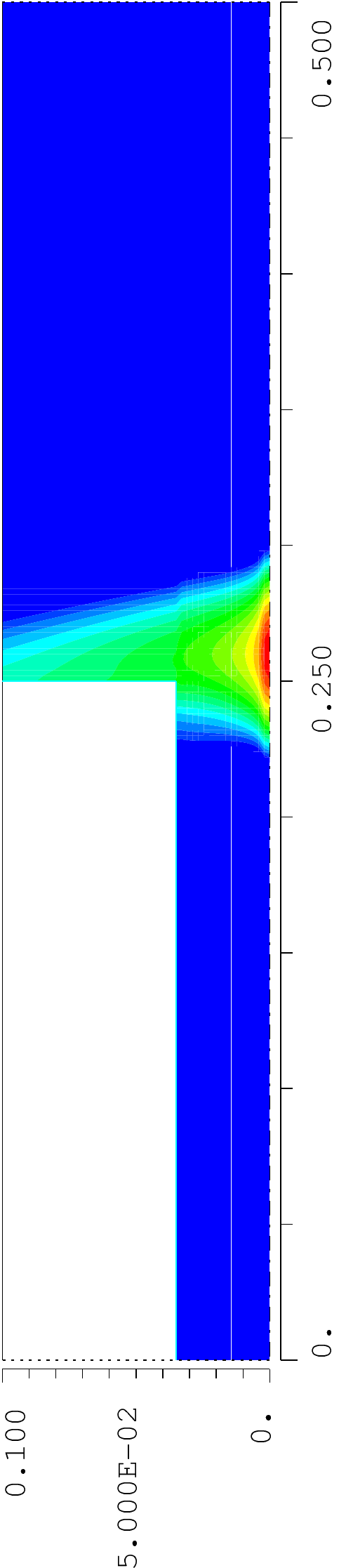}}}    

    \rotatebox{270}{\resizebox{0.15\textwidth}{!}{%
         \includegraphics*[viewport=0 0 142 643,clip]{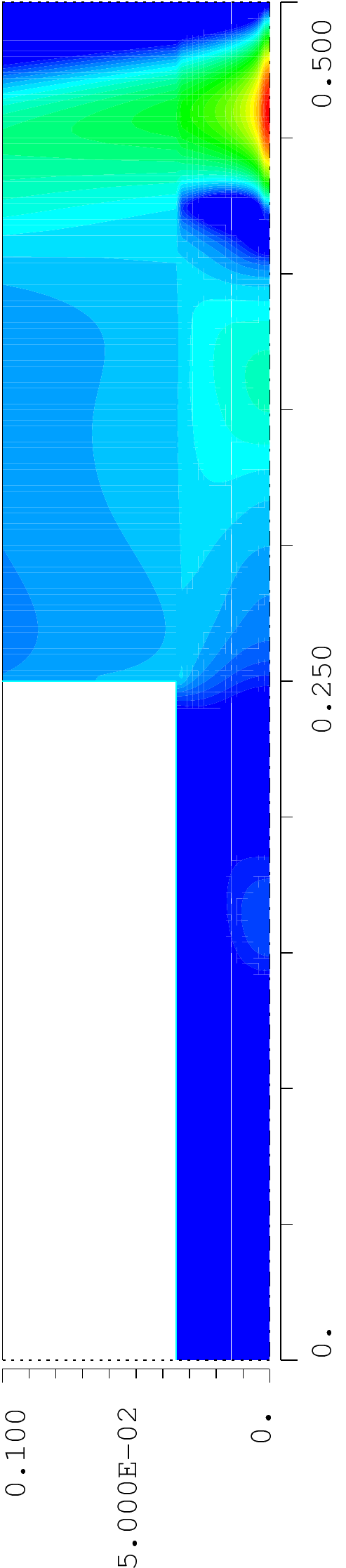}}}
   \end{center}
\caption{
Wake fields behind a bunch generated at a step-out transition from a small
to a larger beam pipe
 }
\label{fig02}
\end{figure}

Consider some typical settings where electromagnetic fields occur
behind a bunch with charge $ q $ moving with velocity $ c $ through a
structure. A bunch moves through a cylindrical pipe along the
$z$-axis. All electric field lines terminate transversely on surface
charges on the wall of the pipe, assuming a perfectly conducting wall.
There will be no wake fields behind the charge. The situation
is different, however, if the cross-section of the beam pipe changes.
A step-out transition is shown in Fig.~\ref{fig02}. All fields have
been calculated using a numerical wake field solver from MAFIA or the
CST studio suite \cite{Weiland80,Weiland84,mafia,CST}.
\setlength{\unitlength}{1cm}
Here we have assumed that all pipe walls are perfect conductors.
The wake field is generated because of the change in geometry.
It should be noted that any beam pipe with finite conductivity, as well as flat beam pipes,
can generate wake fields (resistive wall wake fields) \cite{Piwinski}.
Furthermore, a dielectric-coated pipe, which could
be used as a travelling-wave acceleration section, will generate
wake fields; see, for example, \cite{Steinigke}.

\begin{figure}[thbp]
  \begin{center}
    \rotatebox{270}{\resizebox{2.3cm}{!}{%
        \includegraphics*[bb=230 34 352 790,clip]{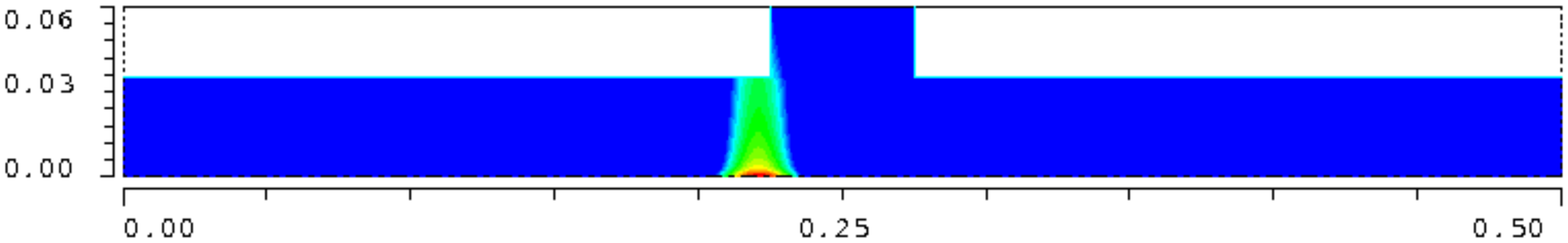}}}
    \rotatebox{270}{\resizebox{2.3cm}{!}{%
        \includegraphics*[bb=230 34 352 790,clip]{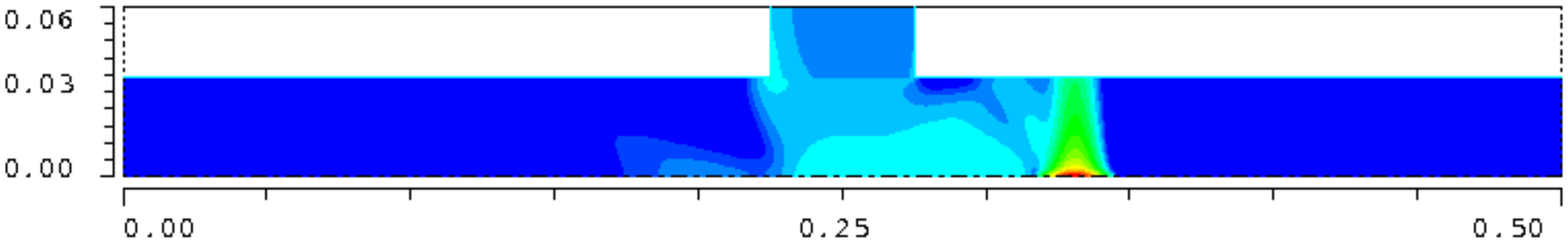}}}
  \end{center}
\caption{
Wake fields in a cavity
 }
\label{fig03}
\end{figure}

Another example is a cavity in a beam pipe; see Fig.~\ref{fig03}.
Again, a bunch is moving through a cylindrical pipe along the $z$-axis.
Wake fields are generated because of geometrical changes in the pipe
cross-section. In this respect the situation is similar to the previously
considered case  of a step-out transition. The main difference is that
the bunch can excite modes in the cavity and therefore long-range wake fields,
which can ring for a long time in the cavity (depending on the conductivity of the cavity wall).
\setlength{\unitlength}{1cm}

The examples have demonstrates that wake forces are caused by geometrical changes along the
beam pipe. Space charge effects are negligible for ultra-relativistic
particles. Wake fields due to the resistive wall or dielectric coatings
should always be checked in detail according to the specific situation.
Furthermore, the surface roughness of the vacuum chamber can be
important for special cases \cite{Dohlus:2001bu}.

\section{Wake fields}
\subsection{Wake fields in a resonant cavity with beam pipes}
The examples above give us a qualitative understanding
of wake fields and how they are generated. Before proceeding to 
mathematical descriptions in terms of wake potentials, let us take a closer
look at the example considered in Section~1.

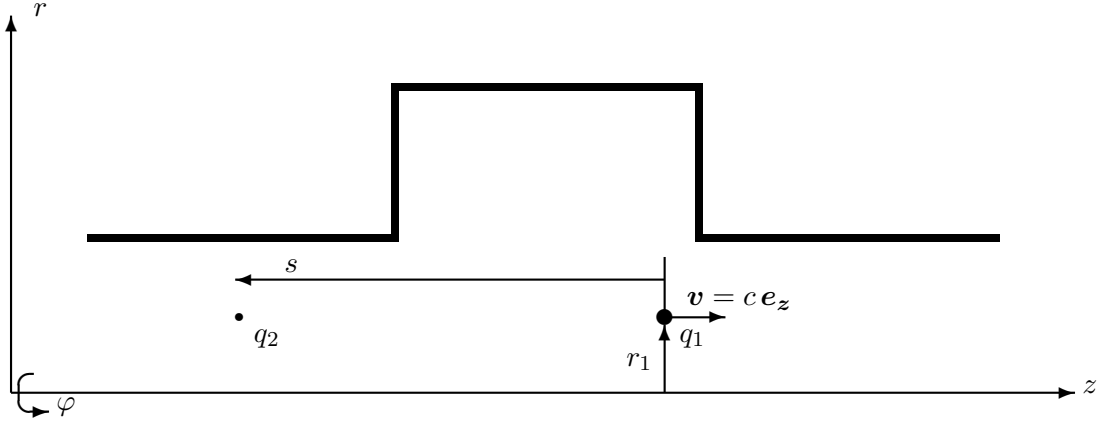
\begin{figure}[hhhhb]
\setlength{\unitlength}{1mm}
\begin{center}
\begin{picture}(150,50)(0,10)
\thicklines
%
\put(2,60){\makebox(0,0)[bl]{ $ r $}}
\put(0,10){\vector(0,1){50} }
\put(140,10){\makebox(0,0)[bl]{ $ z $}}
\put(0,10){\vector(1,0){140} }
\put(3,10){\oval(4,5)[l] }
\put(4,7.5){\vector(1,0){1} }
\put( 5,7.0){\makebox(0,0)[bl]{ $ \varphi $}}
%
\put(50,50){\rule{41mm}{1mm}}
\multiput(50,30)(40,0){2}{\rule{1mm}{20mm} }
\multiput(10,30)(80,0){2}{\rule{40mm}{1mm} }
%
\put(87,16){\makebox(0,0)[bl]{ $ q_1 $}}
\put(86,20){\circle*{2} }
\put(88,21){\makebox(0,0)[bl]{ $ \VEC{v} = c \, \VEC{e_z} $}}
\put(85,20){\vector(1,0){9} }
\put(86,10){\vector(0,1){9} }
\put(80,13){\makebox(0,0)[bl]{ $ r_1 $}}
\put(31,16){\makebox(0,0)[bl]{ $ q_2 $}}
\put(30,20){\circle*{1} }
\put(86,25){\vector(-1,0){56.5} }
\put(86,20){\line(0,1){8} }
\put(35,26){\makebox(0,0)[bl]{ $ s $}}
\end{picture}
\end{center}
\vspace*{-0.3cm}
\caption{An ultra-relativistic point particle with charge $ q_1 $ traverses
a small cavity parallel to the $z$-axis, followed by a test charge~$q_2$
 }
\label{fig04}
\end{figure}

An ultra-relativistic point particle with charge $ q_1 $ traverses
a small cavity parallel to the $z$-axis, with offset $(x_1, y_1)$; see
Fig.~\ref{fig04}.
The electromagnetic force on any test charge $ q_2 $ is given as a function of space and time coordinates by the Lorentz equation 
$$
 \VEC{F}(\VEC{r},t) = q_2 \bigl( \VEC{E}(\VEC{r},t) +
      \VEC{v} \times \VEC{B}(\VEC{r},t) \bigr),
$$
where $\VEC{E}$ and $\VEC{B}$ are the fields generated by $q_1$;
they are solutions of the Maxwell equations
\begin{alignat*}{4}
 \VEC{\nabla} \times \VEC{B} & =  \mu_0 \VEC{j} + \frac{1}{c^2}
                               \frac{\partial}{\partial t} \VEC{E}, &
      \qquad 
\VEC{\nabla} \cdot \VEC{B}  & = 0,\\
 \VEC{\nabla} \times \VEC{E} & =  - \frac{\partial}{\partial t} \VEC{B}, & 
 \qquad
\VEC{\nabla} \cdot \VEC{E} &  =  \frac{1}{\epsilon_0} \rho 
\end{alignat*}
and have to satisfy several boundary conditions.

In our case the charge and current distributions are
$$
\rho(\VEC{r},t) = q_1 \, \delta(x-x_1)\delta(y-y_1)\delta(z-ct),
$$
$$
 \VEC{j}(\VEC{r},t) = c \, \VEC{e_z} \, \rho(\VEC{r},t) .
$$
After interaction of $q_1$ with the cavity, there remain electromagnetic
fields in the cavity. The source particle has lost energy to cavity modes
and excited fields that propagate in the semi-infinite beam pipes.

Now consider a test charge $ q_2$ following $q_1$ at a distance $ s $
with the same velocity $v \approx c $ and with offset $(x_2, y_2)$.
The Lorentz force is
$$
 \VEC{F}(x_1,y_1,x_2,y_2,s,t) = q_2 \bigl( \VEC{E}(x_2,y_2,z=ct-s,t) +
     c\, \VEC{e_z} \times \VEC{B}(x_2,y_2,z=ct-s,t) \bigr).
$$
The change in momentum of the test charge can be calculated as the time-integrated Lorentz force,
$$
  \Delta \VEC{p}(x_1,y_1,x_2,y_2,s) =\int \VEC{F} \,\rd t.
$$
This leads to the concept of wake functions.

The electromagnetic fields $\VEC{E}_{\rm d}$ and $\VEC{B}_{\rm d}$ and the Lorentz
Force $\VEC{F}_{\rm d}$ of a distributed source
 $\rho_{\rm d}(\VEC{r},t) = \eta(x_1-{\bar x}_{1},y_1-{\bar y}_{1})\lambda(z-ct)$
can be calculated either by integration over all source points,
$$
\VEC{F}_{\rm d}({\bar x}_{1},{\bar y}_1,x_2,y_2,s,t) =
\int \VEC{F}(x_1,y_1,x_2,y_2,s,t+z_1/c)
\eta(x_1-{\bar x}_{1},y_1-{\bar y}_{1}) \frac{\lambda(z_1)}{q_1} \,\rd x_1 \,\rd y_1 \,\rd z_1,
$$
or by solving the electromagnetic problem for the distributed source; here 
$\lambda$ is the line charge density, $\eta$ is the transverse distribution
normalized to 1, and ${\bar x}_{1}$ and  ${\bar y}_{1}$ describe a transverse shift
of the center of the distribution. A calculation of the electric fields of a
distributed source is shown in Fig.~\ref{fig05}.

A fundamental difference between fields of point particles
(with time dependency $\delta(z-ct)$) and fields of distributed sources
(with time dependency $\lambda(z-ct)$) is that the frequency spectrum of
point particles is not limited. In particular, long Gaussian bunches may stimulate
only a few resonances (modes) in a cavity structure, or even none.

We can distinguish between the long-range regime of the wake, where the interaction
between particles is driven by resonances, and the short-range regime, where the
superposition of time-harmonic cavity fields is not sufficient to describe the effects.
For instance, the fields in Fig.~\ref{fig02} are not determined by
oscillations, while the fields in Fig.~\ref{fig05} will ring harmonically on one or several
frequencies after the (source) bunch has left the domain.
\setlength{\unitlength}{1.3cm}
\begin{figure}[hbtp]
  \begin{center}
\begin{picture}(7,8)
%
\thicklines
    \put(-0.1,-0.1){\vector(0,1){8} }
    \put(-0.1,-0.1){\vector(1,0){7} }

%
\put(7.0,-0.2){$z$}
\put(-0.3,8.05){$t$}
\put(0,7.1){%
        \includegraphics*[bb=239 34 352 790,clip,angle=270,scale=0.32]{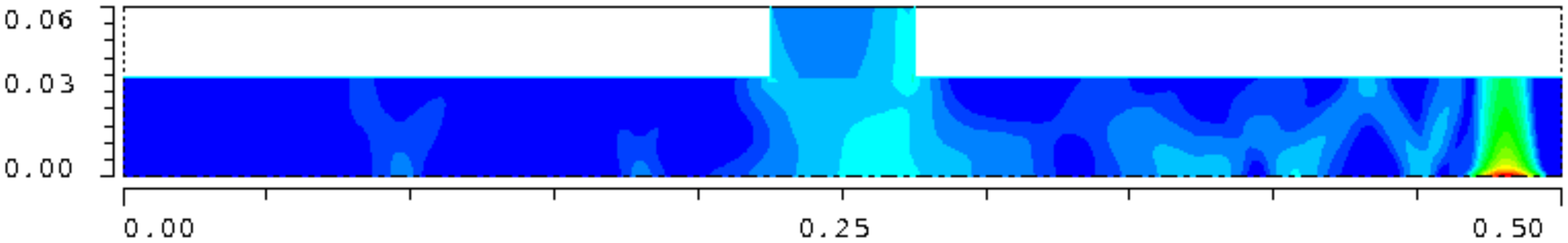}  
        } 
\put(0,6.1){%
        \includegraphics*[bb=239 34 352 790,clip,angle=270,scale=0.32]{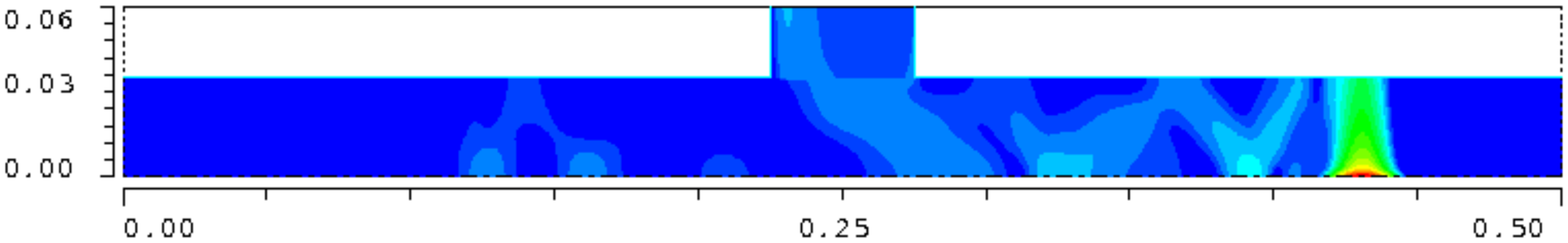}
        }              
\put(0,5.1){%
        \includegraphics*[bb=239 34 352 790,clip,angle=270,scale=0.32]{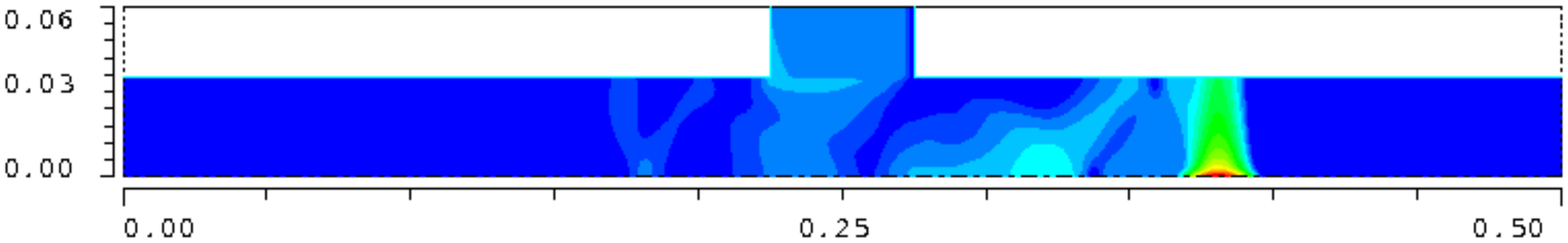}
        }        
\put(0,4.1){%
        \includegraphics*[bb=239 34 352 790,clip,angle=270,scale=0.32]{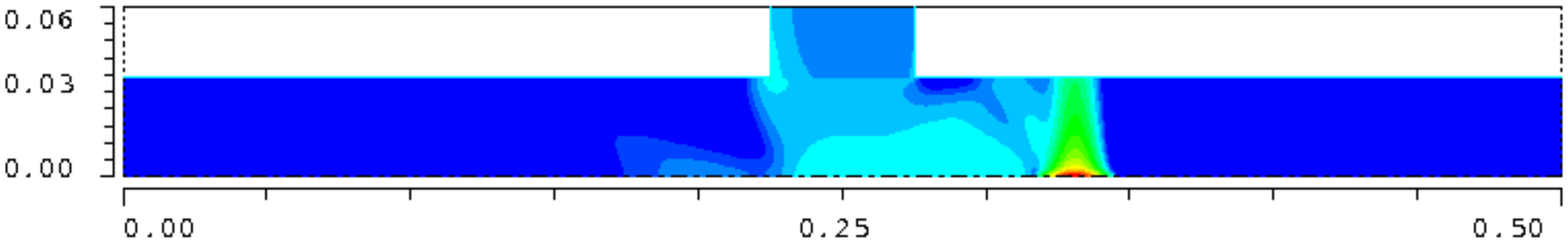}
        }               
\put(0,3.1){%
        \includegraphics*[bb=239 34 352 790,clip,angle=270,scale=0.32]{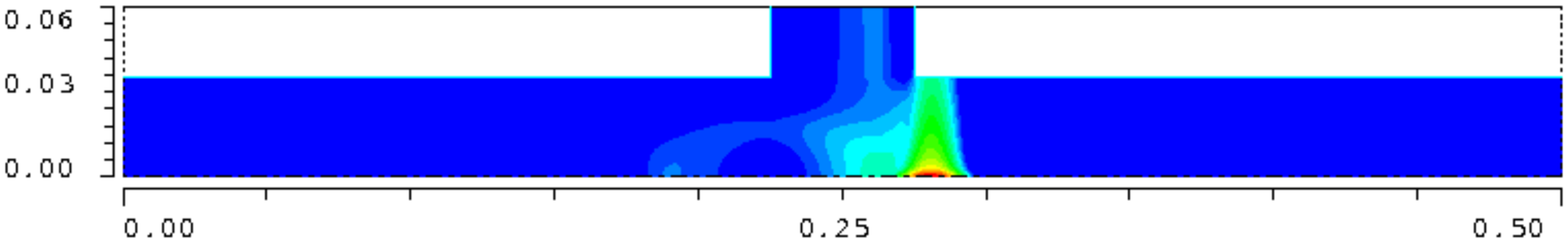}
        }
\put(0,2.1){%
        \includegraphics*[bb=239 34 352 790,clip,angle=270,scale=0.32]{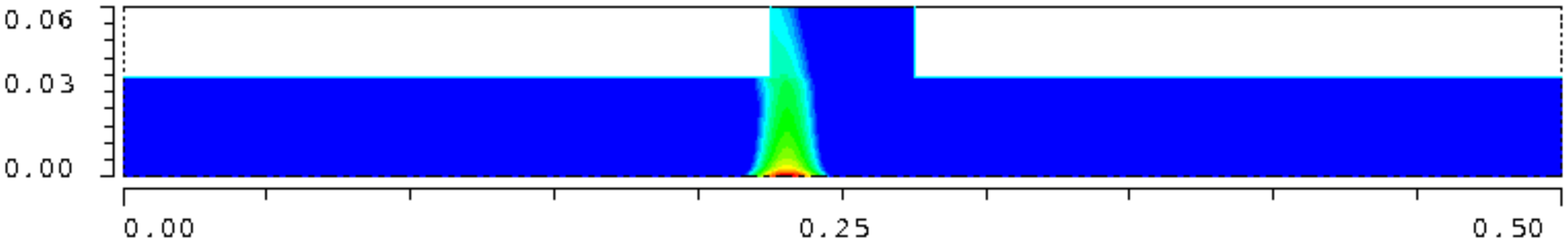}
        }
\put(0,1.0){%
        \includegraphics*[bb=239 34 352 790,clip,angle=270,scale=0.32]{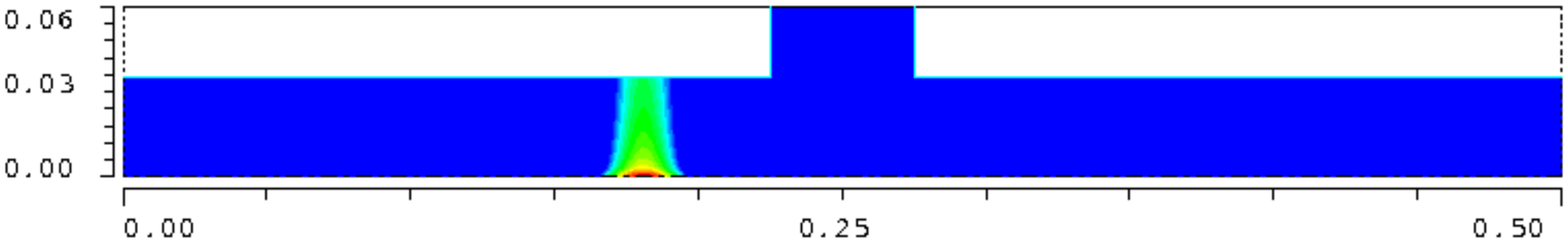}
        }        
    \put(2.7,0.1){\rotatebox{-31}{\line(0,1){8.5}}}
\thinlines
    \put(2.0,0.1){\rotatebox{-31}{\line(0,1){8.5}}}
    \put(7.1,7.4){\vector(-1,0){0.7}}
    \put(6.3,7.5){$s$}
\end{picture}
\end{center}
\caption{
Wake fields generated by a Gaussian bunch traversing a cavity
 }
\label{fig05}
\end{figure}

\subsection{Basic definitions}
Consider a point charge $ q_1 $ traversing a structure with offset
$ (x_1, y_1 )$ parallel to the $z$-axis at the speed of light
(see Fig.~\ref{fig04}).
Then the \emph{wake function} is defined as
$$
 \VEC{w}(x_1,y_1,x_2,y_2,s) = \frac{1}{q_1} \int_{-\infty}^{\infty} \rd z\,
 {\bigl[ \VEC{E}(x_2,y_2,z,t) +  c \, \VEC{e}_z \times
  \VEC{B}(x_2,y_2,z,t) \bigr] }_{t=(s+z)/ c}.
$$
The distance $ s $ is measured from the source $ q_1 $ in the opposite
direction to $ \VEC{v}$.
The change in momentum of a test particle with charge $ q_2 $ following
behind at a distance $ s $ with offset $(x_2, y_2)$ is given by
$$
 \Delta \VEC{p} = \frac{1}{c} \, q_1 q_2 \VEC{w}(s).
$$

Since $ \VEC{e_z} \cdot ( \VEC{e_z} \times \VEC{B} ) = 0 $,
the longitudinal component of the wake function is  simply 
$$
 w_{\|}(x_1,y_1,x_2,y_2,s) = \frac{1}{q_1} \int_{-\infty}^{\infty} \rd z \,
  E_{z}(x_2,y_2,z,(s+z)/c).
$$

\begin{figure}[hbtp]
  \begin{center}
   \rotatebox{0}{\resizebox{10cm}{!}{%
        \includegraphics*{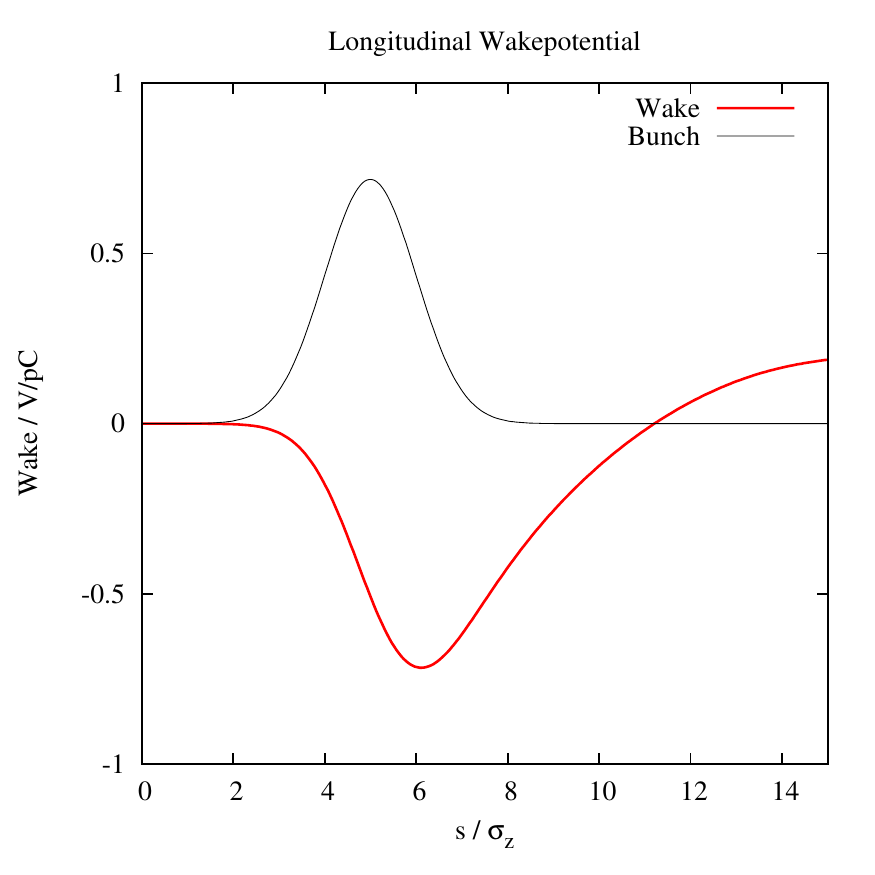}}}
  \end{center}
\caption{
Longitudinal wake potential
 }
\label{fig06}
\end{figure}

Figure \ref{fig06} shows the longitudinal component of the wake potential
for the above example with the small cavity.
The grey line represents the Gaussian charge distribution in the
range from $ - 5 \sigma $ to $ 10\sigma $.
Owing to  transient wake field effects, the head of the bunch
(left-hand side of the figure) is decelerated, while a test charge at
a certain position behind the bunch will be accelerated.
\setlength{\unitlength}{1cm}

The notion of wakes, as presented above, is restricted to sources and
test particles that travel at the velocity of light through a structure with
semi-infinite input and output beam pipes. Therefore, for the
 integrals to converge, it is necessary that there be no length-independent forces in the
pure beam pipes. This is the case for $v \rightarrow c$ and perfect conductivity
of the pipes. The concept of  a wake per length,
$$
 \VEC{w}'(x_1,y_1,x_2,y_2,s) = \frac{1}{q_1}
 \bigl[ \VEC{E}_{\rm p}(x_2,y_2,-s,0) +  v \, \VEC{e}_z \times \VEC{B}_{\rm p}(x_2,y_2,-s,0) \bigr],
$$
is used to describe the effect in beam pipes `p' of finite conductivity and/or
velocity $v \le c$. Suppose that the input and output beam pipes have the same
cross-section; then a generalized wake function
$$
 \VEC{w}_{\rm s}(x_1,y_1,x_2,y_2,s) = \frac{1}{q_1} \int_{-\infty}^{\infty} \rd z\,
 {\bigl[ \VEC{E}_{\rm s}(x_2,y_2,z,t) +  v \, \VEC{e}_z \times
  \VEC{B}_{\rm s}(x_2,y_2,z,t) \bigr] }_{t=(s+z)/ v}
$$
can be defined for the scattered fields $\VEC{E}_{\rm s}=\VEC{E}-\VEC{E}_{\rm p}$ and 
$\VEC{B}_{\rm s}=\VEC{B}-\VEC{B}_{\rm p}$. If the conditions for the wake function
are fulfilled (i.e.\ convergence of the integral), then the wake function equals the
generalized wake function.

The \emph{wake potential} is defined similarly to the wake function, but for a distributed source:
\begin{align*}
 \VEC{W}({\bar x}_1,{\bar y}_1,x_2,y_2,s) &= \frac{1}{q_1} \int_{-\infty}^{\infty} \rd z\,
 {\bigl[ \VEC{E}_{\rm d}(x_2,y_2,z,t) +  c \, \VEC{e}_z \times
  \VEC{B}_{\rm d}(x_2,y_2,z,t) \bigr] }_{t=(s+z)/ c}
\nonumber \\
   &= \frac{1}{q_1 q_2} \int_{-\infty}^{\infty} \rd z\,
 {\bigl[ \VEC{F}_{\rm d}({\bar x}_1,{\bar y}_1,x_2,y_2,z,t) \bigr] }_{t=(s+z)/ c}.
\end{align*}
It can be calculated from the wake function by the convolution
$$
\VEC{W}_{\rm d}({\bar x}_{1},{\bar y}_1,x_2,y_2,s) =
\int \VEC{w}(x_1,y_1,x_2,y_2,s+z_1)
\eta(x_1-{\bar x}_{1},y_1-{\bar y}_{1}) \frac{\lambda(z_1)}{q_1} \,\rd x_1 \,\rd y_1 \,\rd z_1.
$$
Note that the $s$-coordinate measures in the negative $z$-direction while $\lambda$ depends
on the positive longitudinal coordinate. Usually numerical codes for computing wakes, such as ECHO,
calculate electromagnetic fields for distributed sources and therefore wake
potentials.

\subsection{Some theory}
\subsubsection{The Panofsky--Wenzel theorem}

We follow the arguments of A.\ Chao \cite{Chao_CAS2015, Chao93} to introduce the
Panofsky--Wenzel theorem \cite{Panofsky}. Therefore we use the following different notation for the generalized wake function:
$$
\VEC{w}_{\rm p}(x_1,y_1,x_2,y_2,s)=\VEC{w}_{\rm p}(x_1,y_1,\VEC{r}_2),
$$
with the observer vector $\VEC{r}_2=x_2 \VEC{e}_x+y_2 \VEC{e}_y-s \VEC{e}_z$.
We calculate curl $\VEC{w}_{\rm p}$ with respect to the observer or the test particle:
$$
\nabla_2 \times \VEC{w}_{\rm p}(x_1,y_1,\VEC{r}_2)=
\nabla_2 \times
 \frac{v}{q_1} \int_{-\infty}^{\infty} \rd t\,
 {\bigl[ \VEC{E}_{\rm s}(\VEC{r}_2+\VEC{v}t,t) +  \VEC{v} \times
  \VEC{B}_{\rm s}(\VEC{r}_2+\VEC{v}t,t) \bigr] }.
$$
Using ${\rm curl}\,\VEC{E}=-\partial \VEC{B}/\partial t$ gives
\begin{align*}
\nabla_2 \times \VEC{w}_{\rm p}(x_1,y_1,\VEC{r}_2)&=
 \frac{v}{q_1} \int_{-\infty}^{\infty} \rd t\,
 {\left[ -\frac{\partial}{\partial t}\VEC{B}_{\rm s}(\ldots ,t)
         +  \VEC{v} (\nabla_2 \VEC{B}_{\rm s}(\ldots ,t))
         -  \VEC{B}_{\rm s}(\ldots ,t) (\nabla_2 \VEC{v}) \right] }
\nonumber \\
&= \frac{v}{q_1} \int_{-\infty}^{\infty} \rd t\,
   {\left[ -\frac{\partial}{\partial t}-v\frac{\partial}{\partial z}
    \right] \VEC{B}_{\rm s}(\VEC{r}_2+\VEC{v}t,t)}
\nonumber \\
&= \frac{v}{q_1} \int_{-\infty}^{\infty} \rd t\,
    {\left[ -\frac{\rd}{\rd t} \VEC{B}_{\rm s}(\VEC{r}_2+\VEC{v}t,t) \right] }
\nonumber \\
&= -\frac{v}{q_1} \VEC{B}_{\rm s}(\VEC{r}_2+\VEC{v}t,t) \Big|_{t=-\infty}^{t=\infty}.
\nonumber
\end{align*}
As the scattered field is zero for negative infinite time and vanishes for
positive infinite time and infinite distance from the scattering object,
the wake is curl-free. The Panofsky--Wenzel theorem is then reformulated in our
original notation as the set of equations
\begin{align*}
    \frac{\partial}{\partial s  } w_{{\rm p}x}     
    &= -
    \frac{\partial}{\partial x_2} w_{{\rm p} \|},  
\nonumber \\
    \frac{\partial}{\partial s  } w_{{\rm p}y}     
    &= -
    \frac{\partial}{\partial y_2} w_{{\rm p} \|},  
\nonumber \\
    \frac{\partial}{\partial x_2} w_{{\rm p}y}     
    &=
    \frac{\partial}{\partial y_2} w_{{\rm p}x}.    
\nonumber
\end{align*}
Note that the Panofsky--Wenzel theorem holds for the generalized wake
function ($v \le c$) and for the wake function ($v=c$).

Integration of the transverse gradient of the longitudinal wake function
yields the transverse wake potential
$$
 \VEC{w}_{\bot}(x_1,y_1,x_2,y_2,s)
= - \VEC{\nabla}_{2 \bot} \int_{-\infty}^{s}\rd s' \, w_{\|} (x_1,y_1,x_2,y_2,s').
$$
The Panofsky--Wenzel theorem is applicable if the input and
output beam pipes have the same cross-section.
\subsubsection{Wake is harmonic with respect to observer offset}
Now we calculate div$\,\VEC{w}$ with respect to the observer. First, note that
$$
\nabla_2 \cdot \VEC{w}(x_1,y_1,\VEC{r}_2)=
\nabla_2 \cdot
 \frac{c}{q_1} \int_{-\infty}^{\infty} \rd t\,
  {\bigl[ \VEC{E}(\VEC{r}_2+\VEC{c}t,t) +  \VEC{c} \times
  \VEC{B}(\VEC{r}_2+\VEC{c}t,t) \bigr] }.
$$
Using Maxwell's equations, ${\rm div}\,\VEC{E} = \rho/\varepsilon$ and
${\rm curl}\,\VEC{B}=\mu \VEC{J} +c^{-2} \partial \VEC{E}/\partial t$,
together with $\VEC{J} = \VEC{c} \rho$ gives
\begin{align*}
\nabla_2 \cdot \VEC{w}(x_1,y_1,\VEC{r}_2)&=
 \frac{c}{q_1} \int_{-\infty}^{\infty} \rd t\,
 {\bigl[ \nabla_2 \cdot \VEC{E} +  \VEC{c} (\nabla_2 \times \VEC{B}) \bigr] }
\nonumber \\
  &=
 \frac{1}{q_1} \int_{-\infty}^{\infty} \rd t\,
 {\left[ -\frac{\partial}{\partial t} E_z(\VEC{r}_2+\VEC{c}t,t) \right] }
\nonumber \\
  &=
 -\frac{1}{q_1} \int_{-\infty}^{\infty} \rd z\,
 {\left[ \frac{\partial}{\partial s} E_z(\VEC{r}_2+z\VEC{e}_z,(z+s)/c) \right] }
\nonumber \\
  &=
 -\frac{\partial}{\partial s} \,w_\|(x_1,y_1,x_2,y_2,s)
.
\nonumber
\end{align*}
The term $\partial w_\| / \partial s$ appears on both sides of the equation, so
we can write
$$
\frac{\partial w_x}{\partial x_2} + \frac{\partial w_y}{\partial y_2}=0
.
$$
With the Panofsky--Wenzel equations we find that the longitudinal wake is a
harmonic function with respect to the transverse coordinates of the test particle:
$$
\left(\frac{\partial^2}{\partial x_2^2}+\frac{\partial^2}{\partial y_2^2} \right) w_\|=
-\frac{\partial}{\partial s} \left( \frac{\partial w_x}{\partial x_2} + \frac{\partial w_y}{\partial y_2} \right)=
0.
$$
 It is required that the trajectories $(x_1,y_1,ct)$ and $(x_2,y_2,ct)$ do not intersect
with the boundary.
\subsubsection{Wake is harmonic with respect to source offset}
The longitudinal wake is also a harmonic function with respect to the transverse
coordinates of the source particle \cite{Zagorodnov}, i.e.\ $L_1 w_\|=0$ with
$L_1=\partial^2/\partial x_1^2 + \partial^2/\partial y_1^2$. To prove this, we
have to calculate $\tilde{E}_z = L_1 E_z$,  which is equivalent to the solution
of the field problem for the source $\tilde{\rho} = L_1 \rho$.
The source $\rho$ is the point particle $q_1$ travelling
at the speed of light along $(x_1,y_1,z=ct)$. It gives rise to the electromagnetic fields
\begin{align*}
\VEC{E}_{\rm f} &= q_1 \frac{\delta(z-ct)}{2 \pi \varepsilon}
                 \frac{(x-x_1)\VEC{e}_x+(y-y_1)\VEC{e}_y}{(x-x_1)^2+(y-y_1)^2},
\\ \nonumber
\VEC{B}_{\rm f} &= c^{-1}  \VEC{e}_z \times \VEC{E}_{\rm f}
\nonumber
\end{align*}
in free space. The fields $\tilde{\VEC{E}}=L_1 \VEC{E}_{\rm f}$ and
$\tilde{\VEC{B}}=L_1 \VEC{B}_{\rm f}$ are caused by the source $\tilde{\rho}=L_1 \rho$.
These fields are zero for all points with $(x,y)\ne (x_1,y_1)$, as
$$
\left( \frac{\partial^2}{\partial x_1^2}+\frac{\partial^2}{\partial y_1^2} \right)
\frac{(x-x_1)\VEC{e}_x+(y-y_1)\VEC{e}_y}{(x-x_1)^2+(y-y_1)^2}=\VEC{0}.
$$
Obviously $\tilde{\VEC{E}}$ and $\tilde{\VEC{B}}$ satisfy
any linear boundary condition for any geometry, provided that the boundary does not intersect the trajectory
$(x_1,y_1,z=ct)$. Therefore these fields are also solutions to the bounded wake
problem, and all components of $\VEC{w}$ are harmonic with respect to $(x_1,y_1)$, since
$$
\left( \frac{\partial^2}{\partial x_1^2}+\frac{\partial^2}{\partial y_1^2} \right) \VEC{w}
=
 \frac{c}{q_1} \int_{-\infty}^{\infty} \rd t\,
 {\bigl[\tilde{\VEC{E}} +  \VEC{c} \times \tilde{\VEC{B}} \bigr] }=\VEC{0}
.
$$


This information will help us to evaluate the $ r $-dependence
of the wake function in cylindrical symmetric structures
in the next subsection, and it will enable us to  efficiently calculate the wake function
in fully 3D structures.

\subsection{Wake function in cylindrically symmetric structures}
\begin{figure}[hbtp]
\setlength{\unitlength}{1mm}
\begin{center}
\begin{picture}(150,60)(0,10)
\thicklines
%
\put(2,60){\makebox(0,0)[bl]{ $ r $}}
\put(0,10){\vector(0,1){50} }
\put(140,10){\makebox(0,0)[bl]{ $ z $}}
\put(0,10){\vector(1,0){140} }
\put(3,10){\oval(4,5)[l] }
\put(4,7.5){\vector(1,0){1} }
\put( 5,7.0){\makebox(0,0)[bl]{ $ \varphi $}}
%
\put(120,10){\vector(0,1){20} }
\put(121,20){\makebox(0,0)[bl]{ $ a $}}
\put(49,49.5){\rule{43mm}{0.5mm}}
\multiput(10,30)(79,0){2}{\rule{42mm}{0.5mm} }
%
\multiput(50,35)(0.25,0){4}{\oval(10,10)[r]}
\multiput(50,45)(0.25,0){4}{\oval(10,10)[l]}
\multiput(90,35)(0.25,0){4}{\oval(10,10)[l]}
\multiput(90,45)(0.25,0){4}{\oval(10,10)[r]}
%
\put(79,22){\makebox(0,0)[bl]{ $ q_1 $}}
\put(85,20){\circle*{2} }
\put(88,21){\makebox(0,0)[bl]{ $ \VEC{v} = c \, \VEC{e_z} $}}
\put(85,20){\vector(1,0){9} }
\put(85,20){\line(-1,0){10} }
\put(80,10){\vector(0,1){10} }
\put(81,15){\makebox(0,0)[bl]{ $ r_1 $}}
\put(24,17){\makebox(0,0)[bl]{ $ q_2 $}}
\put(30,15){\circle*{2} }
\put(30,15){\line(1,0){10} }
\put(35,10){\vector(0,1){5} }
\put(36,11){\makebox(0,0)[bl]{ $ r_2 $}}
\put(85,25){\vector(-1,0){55} }
\put(85,20){\line(0,1){8} }
\put(30,15){\line(0,1){13} }
\put(35,26){\makebox(0,0)[bl]{ $ s $}}
\end{picture}
\end{center}
\caption {
A bunch with total charge $ q_1 $ traversing a cavity with offset $ r_1 $,
followed
by a test charge $ q_2 $ with offset $ r_2 $
}
\label{fig14}
\end{figure}
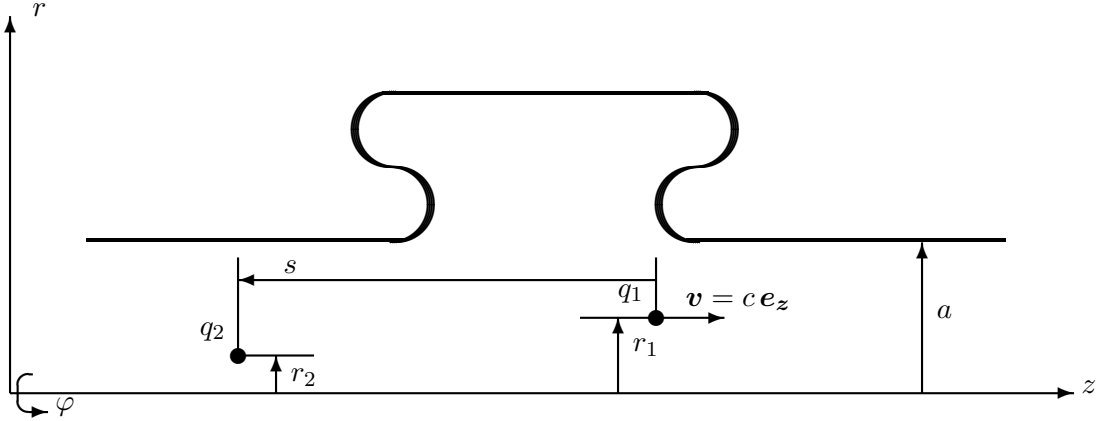
Consider now a cylindrically symmetric acceleration cavity with side tubes
of radius $ a $ (see Fig.~\ref{fig14}). The particular shape in the
region $ r > a $ is of no importance for the following investigations.
Two charges pass through the structure from left to right with the speed of
light: $ q_1 $ at a radius of $ r_1 $ and $ q_2 $ at a radius of $ r_2 $.
We wish to find an expression for the net change in momentum,
$ \Delta \VEC{p} (r_1,\varphi_1,r_2,\varphi_2,s) $, experienced by $ q_2 $
due to the wake fields generated by $ q_1 $. In the following we
write the wake function and potential in polar coordinates. Let
us start with the case of $ \varphi_1 = 0 $:
$$
 \Delta p_z (r_1,0,r_2,\varphi_2,s) = q_1 \, q_2 \, w_{\|}(r_1,0,r_2,\varphi_2,s).
$$
The wake function can be expanded in a multipole series
$$
w_{\|}(r_1,0,r_2,\varphi_2,s) = {\rm Re} \left\{
\sum_{m=-\infty}^{\infty} \exp({\ri\, m \, \varphi_2}) G_m(r_1,r_2,s)
           \right\}.
$$
Since  $ w_{\|}$ is a harmonic function in $ (r_2, \varphi_2) $, we have
\begin{align*}
L_2 \,w_{\|}(r_1,0,r_2,\varphi_2,s)  & = 
 \left( \frac{1}{r_2} \frac{\partial}{\partial r_2} r_2 \frac{\partial}{\partial r_2}
+ \frac{1}{r^2} \frac{\partial^2 }{\partial \varphi_2^2 } \right)
w_{\|}(r_1,0,r_2,\varphi_2,s)
\nonumber \\
 & =  {\rm Re} \left\{
\sum_{m=-\infty}^{\infty} \exp({\ri\, m \, \varphi_2})
\left( \frac{1}{r_2} \frac{\partial}{\partial r_2} r_2 \frac{\partial}{\partial r_2}
- \frac{m^2}{r_2^2}  \right)
G_m(r_1,r_2,s)  \right\}
\nonumber \\
& =  0,
\nonumber
\end{align*}
where $L_2$ is the transverse Laplace operator with respect to the offset of the test
particle. So, for all $ m $, the expansion functions $ G_m(r_1,r_2,s) $ have to satisfy
the Poisson equation
$$
 \frac{1}{r_2} \frac{\partial}{\partial r_2}
\left( r_2 \frac{\partial}{\partial r_2} G_m(r_1,r_2,s) \right) 
-  \frac{m^2}{r_2^2} G_m(r_1,r_2,s)  =  0.
$$
The solutions are
\begin{align*}
 G_0(r_1,r_2,s) & =  U_0(r_1,s) + V_0(r_1,s) \, \ln r_2,
\nonumber \\
 G_m(r_1,r_2,s) & =  U_m(r_1,s) \, r_2^m + V_m(r_1,s) \, r_2^{-m} \quad
\text{for } m > 0.
\nonumber
\end{align*}

Keeping only the  solutions  which are regular at the origin ($ r_2 = 0$),
the longitudinal wake potential can be written as
$$
w_{\|}(r_1,0,r_2,\varphi_2,s) =
\sum_{m=0}^{\infty} \, r_2^m \, U_m(r_1,s) \, \cos m\varphi_2,
$$
with expansion functions $ U_m(r_1,s) $ that depend on the details of
the given cavity geometry.

By azimuthal symmetry, the dependence on $\varphi_1$ is
$w_{\|}(r_1,\varphi_1,r_2,\varphi_2,s)=w_{\|}(r_1,0,r_2,\varphi_2-\varphi_2,s)$,
as longitudinal fields depend only on the relative azimuthal angle of the
observer with respect to the source. Using the fact that $w_\|$ is also a
harmonic function in $(r_1, \varphi_1)$, we find with the same arguments as
before that $U_m(r_1,s)$ can be factorized as $r_1^m w_m(s)$.

It follows that for the general case of a charge $ q_1 $ at $( r_1, \varphi_1 )$
generating fields that act on a second charge $ q_2 $ at $( r_2, \varphi_2 )$,
the \emph{longitudinal wake function} is given by
$$
w_{\|}(r_1,\varphi_1,r_2,\varphi_2,s) =
\sum_{m=0}^{\infty} r_1^m \, r_2^m \, w_m(s)
 \, \cos m( \varphi_2 - \varphi_1) .
$$
The transverse wake function is, by the Panofsky--Wenzel
theorem,
\begin{align*}
 \VEC{w}_{\bot}(r_1,\varphi_1,r_2,\varphi_2,s) & = 
 - \left( \VEC{e_r} \, \frac{\partial}{\partial r_2} +
     \VEC{e_{\varphi}} \,
     \frac{1}{r_2} \frac{\partial}{\partial \varphi_2} \right)
\int_{-\infty}^{s}\rd s' \, w_{\|} (r_1,\varphi_1,r_2,\varphi_2,s')
\nonumber \\
 & =  \sum_{m=0}^{\infty}  \Biggl\{
-\VEC{e_r} \, m  \,{r_1}^m \, {r_2}^{m-1}
 \int_{-\infty}^{s}\rd s'  \,w_m(s')
 \, \cos m( \varphi_2 - \varphi_1) 
\nonumber\\
 & \qquad \quad 
+\VEC{e_{\varphi}} \, m  \,{r_1}^m \, {r_2}^{m-1}
 \int_{-\infty}^{s}\rd s' \,w_m(s')
 \, \sin m( \varphi_2 - \varphi_1) \Biggr\}.
\nonumber
\end{align*}

Each azimuthal order is fully characterized by a scalar function $w_m(s)$.
This function can be calculated by solving Maxwell's equations for the given
geometry and any choice of $(r_1, \varphi_1, r_2, \varphi_2)$, yielding 
$$
w_m(s) = \frac{ \int_{-\infty}^{\infty} \rd z \, E_{zm}(r_2,\varphi_2,z,(z+s)/c))}
        {r_1^m \, r_2^m \, \cos m (\varphi_2 - \varphi_1)}.
$$
A particular choice of $r_2$ can be used to avoid the infinite integration range:
since $ E_z $ vanishes at the metallic tube boundary, only the cavity gap contributes
to the integral. The integration range is reduced to the cavity gap by setting
$r_2$ to the radius of the beam tube. This trick is possible if no obstacle intersects
with the infinite cylindrical beam pipe.

This type of wake integration is utilized by computer codes such as ECHO \cite{Zagorodnov:2005my,Echo} for bunches of finite length. Wake potentials can be calculated by such programs in the time domain,
but wake functions (of point sources) need asymptotic considerations; see \cite{Zagorodnov2003}.

It should be mentioned that in many practical cases, due to the
$ {(r/a)}^m $ dependence, the longitudinal wake is dominated by the
monopole term and the transverse wakes by the dipole term:
\begin{align*}
w_{\|}(r_1,\varphi_1,r_2,\varphi_2,s)  & =  w_0(s),                    
\nonumber \\
\VEC{w}_{\bot}(r_1,\varphi_1,r_2,\varphi_2,s) & = 
 r_1 \, \int_{-\infty}^{s}\rd s' \,w_1(s') \,
\bigl[
 -\VEC{e_r}  \, \cos ( \varphi_2 - \varphi_1)+\VEC{e_{\varphi}} \, \sin ( \varphi_2 - \varphi_1)
\bigr].
\nonumber
\end{align*}

\subsection{Fully 3D structures}
\begin{figure}[hhhhhtp]
  \begin{center}
  \resizebox{7.5cm}{!}{\includegraphics*{%
        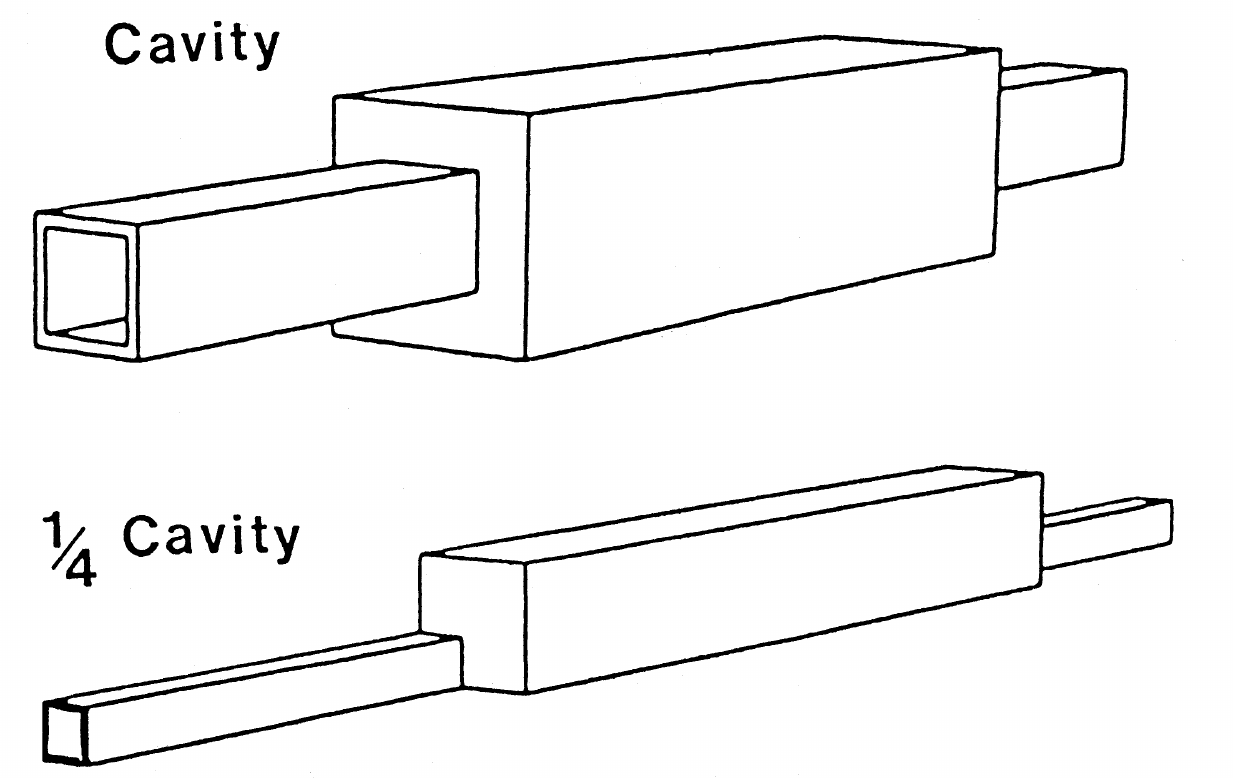}}
  \end{center}
\caption{
A 3D cavity structure with two symmetry planes (top) and a quarter of the structure (bottom)
 }
\label{fig08}
\end{figure}

While for cylindrical symmetric structures the dependence of the wake on
transverse coordinates is explicitly known and can be used to reduce the
integration range and domain of the field calculation, more general structures
require us to use the harmonic property of the wake for a beam tube of
arbitrary shape. The simple 3D cavity in Fig.~\ref{fig08}, with a beam
tube of square cross-section, is used to demonstrate this. We suppress
the dependence of the wake function (or potential) on the offset of the
source and write simply $ \tilde{W}_{\|}(x,y,s)=W_{\|}(x_1,y_1,x,y,s) $. This
function is harmonic in the observer offset,
$$
   \VEC{\nabla}_{\bot}^2 \,\tilde{W}_{\|} (x,y,s) = 0.
$$
\setlength{\unitlength}{1mm}

\begin{figure}[hhhhhtp]
  \begin{center}
$
\begin{array}{cc}
  \resizebox{4cm}{!}{\includegraphics*{%
        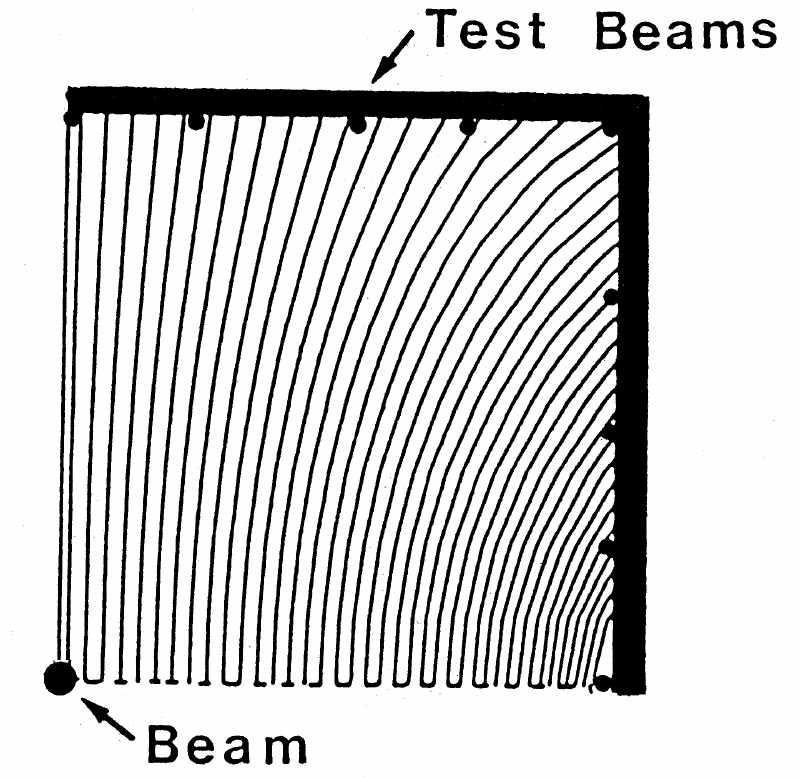}} &
 \resizebox{4cm}{!}{\includegraphics*{%
        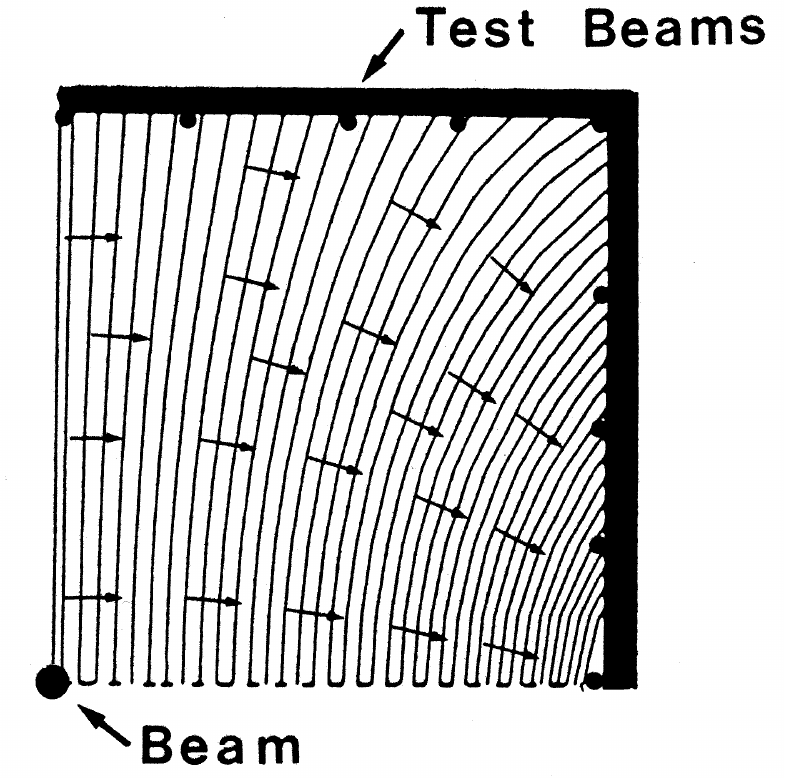}}
\end{array}$
  \resizebox{8cm}{!}{\includegraphics*{%
        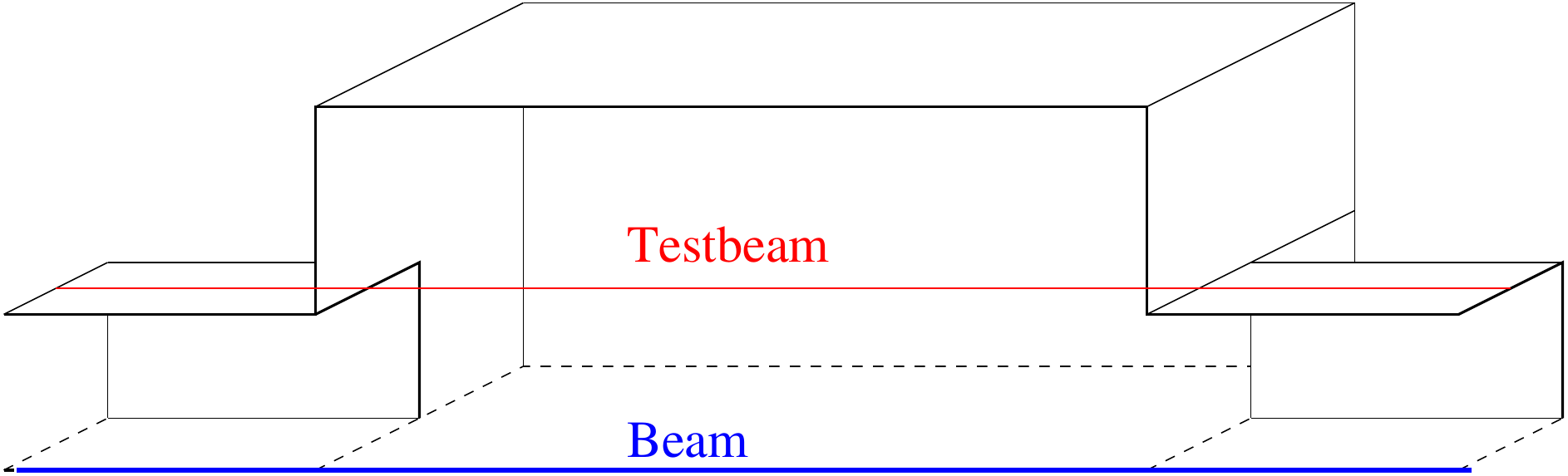}}
  \end{center}
\caption{
Illustration of the indirect test beam method. The upper pictures show lines
of constant longitudinal wake potential and the gradient of the longitudinal wake potential;
an integration gives the transverse wake potential according
to the Panofsky--Wenzel theorem. The lower diagram depicts the paths of the beam and
the test beam.
 }
\label{fig09}
\end{figure}

For points $x$ and $y$ on the surface of the beam tube, we can calculate
the wake by a finite-range integration through the cavity gap, as shown in Fig.~\ref{fig09}.
If we know $\tilde{W}$ for all surface points, we can calculate the wake
for any point inside the tube by numerical solution of the boundary
value problem. Therefore a 2D Poisson problem has to be solved. In our
example, with two transverse symmetries, only a quarter of the structure
needs to be considered to calculate the wake of a source in the center.

\setlength{\unitlength}{1mm}

The transverse wake potential can be calculated from the longitudinal
one using the Panofsky--Wenzel theorem. The transverse gradient of the
longitudinal wake potential in a beam tube is also indicated in Fig.~\ref{fig09}.

\section{Cavities, resonant structures and eigenmodes}
\subsection{Eigenmodes}
Many structures in an accelerator environment can be considered as a
hollow space with semi-infinite beam pipes on both sides. Usually this
vacuum volume is bounded by metal surfaces with high conductivity. As a 
good approximation, the cavity walls can be treated as perfect electric
conducting (PEC) boundaries, and sometimes the beam pipes are even neglected
so that the volume is closed.

Electromagnetic fields with frequencies below the lowest cutoff
frequency of the beam pipes are trapped in the volume, and the 
fields oscillate at discrete frequencies:
\begin{align*}
   \VEC{E}(\VEC{r},t)&=\sum_\nu \hat{A}_\nu \hat{\VEC{E}}_\nu(\VEC{r}) \cos(\hat{\omega}_\nu t+\hat{\varphi}_\nu)
,
\nonumber \\
   \VEC{B}(\VEC{r},t)&=\sum_\nu \hat{A}_\nu \hat{\VEC{B}}_\nu(\VEC{r}) \sin(\hat{\omega}_\nu t+\hat{\varphi}_\nu)
.
\nonumber
\end{align*}
These oscillations are called eigenmodes or cavity modes. They are characterized
by their field patterns $\hat{\VEC{E}}_\nu(\VEC{r})$ and $\hat{\VEC{B}}_\nu(\VEC{r})$ and  their
eigenfrequencies $\hat{\omega}_\nu$. The modes may ring with any amplitude $\hat{A}_\nu$
and phase $\hat{\varphi}_\nu$, and the amplitude normalization of the eigenfields is
arbitrary. Such modes are called standing-wave modes, as the electric and 
magnetic fields ring at all spatial points with the same phase, but the electric
field is  phase-shifted by $90^\circ$ relative  to the magnetic field.
For simplicity, in the following we omit the mode index $\nu$ but indicate
all indexable (mode-specific) quantities with a hat. We will introduce further  
mode-specific quantities, such as the quality $\hat{Q}$, the modal longitudinal loss parameter
$\hat{k}$, and the mode energy
$$
\hat{\cal W}=\frac{1}{2}\int{\varepsilon {\hat E}^2} \,\rd V
         =\frac{1}{2}\int{\mu^{-1} {\hat B}^2} \,\rd V,
$$
which depends on the arbitrary amplitude normalization. The total electromagnetic
field energy of all the modes is\footnote{The field energy of a particular mode
does not depend on the stimulation of other modes, as the mode fields are orthogonal to
each other; see Appendix~A.}
$$
{\cal W}=\frac{1}{2}\int{\varepsilon E(\VEC{r,t})^2} \,\rd V+\frac{1}{2}\int{\mu^{-1} B(\VEC{r,t})^2} \,\rd V
 =\sum |\hat A|^2 \,\hat{\cal W}.
$$

\begin{figure}[hhhhhtp]
  \begin{center}
    \rotatebox{270}{\resizebox{4cm}{!}{%
        \includegraphics*[bb=0 0 224 756,clip]{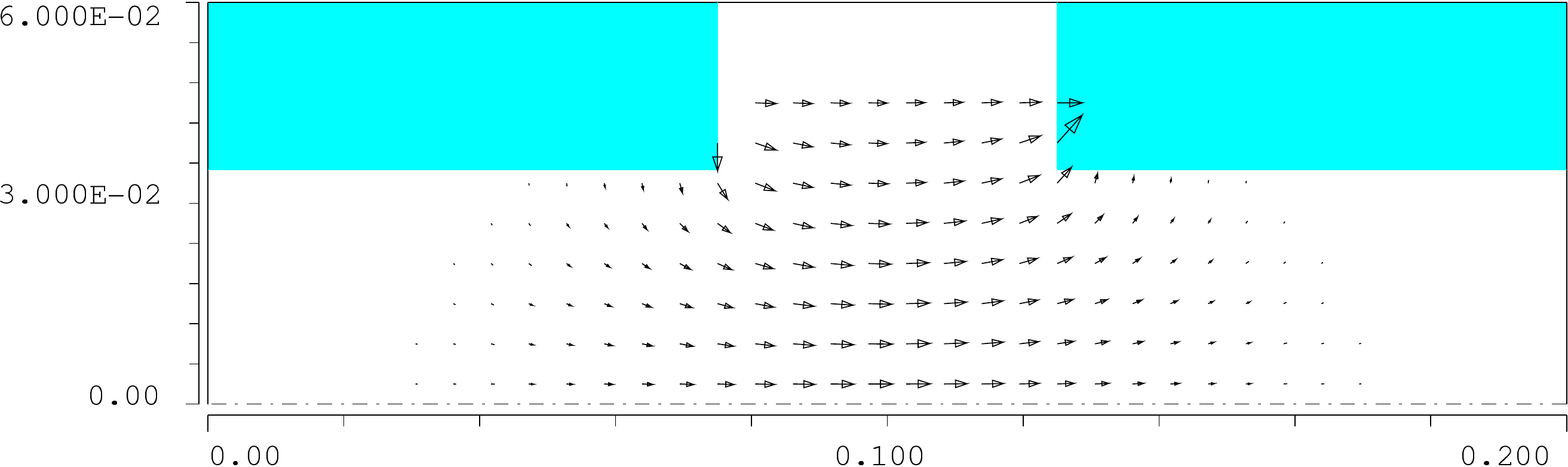}}}
  \end{center}
\caption{
Electric field of a mode in a rotationally symmetric cavity with beam pipes
 }
\label{fig10}
\end{figure}

Eigenmodes can be computed with electromagnetic field solvers such as those in \cite{mafia,CST}; see
also Fig.~\ref{fig10}. Usually closed volumes are considered, which are completely surrounded by PEC or perfect magnetic conducting (PMC) surfaces. As the mode field
in beam pipes decays exponentially, even open problems (involving infinitely long pipes)
can be handled with such programs, by using a perfectly conducting boundary after
a sufficiently long piece of pipe.

\setlength{\unitlength}{1mm}

In structures with symmetries (e.g.\ rotational symmetry), eigenmodes
and beam-pipe modes of the same symmetry condition are coupled. Therefore the
lowest cutoff frequency for a particular symmetry defines the highest possible
eigenfrequency for the corresponding eigenmodes. For instance, monopole modes
may have resonance frequencies that are above the lowest dipole mode cutoff frequency,
which is lower than the lowest monopole mode cutoff frequency.
Beyond that, there can exist quasi-trapped modes above the lowest cutoff frequency
that have very weak coupling to the pipes. The energy flow (per period) of such fields
into the beam pipes may be comparable to the energy loss (per period) of non-trapped
modes to non-perfectly conducting metallic boundaries.

\subsection{Excitation of eigenmodes and the per-mode loss parameter}
\label{section_Excitation_of_Eigenmodes}
We consider a cavity of length\footnote{The relevant length is not exactly the length of the
cavity, but rather the length with non-zero field of the modes. For open structures, with beam pipes,
this length is in principle infinite, but for practical considerations the field has decayed
sufficiently after a pipe length of a few times the widest dimension of the cross-section.}
$L$ and a bunch with charge $q_1$, offset $(x_1, y_1)$ and 
velocity $c$, which enters the cavity at time $t=0$. The electromagnetic fields after
the charge has left the cavity, namely
\begin{align*}
   \VEC{E}(\VEC{r},t>L/c)&=
    \sum {\rm Re} \bigl\{ \hat{A} \hat{\VEC{E}}(\VEC{r}) \exp(\ri\, \hat{\omega} t) \bigr\}
   +\VEC{E}_{\rm r}(\VEC{r},t)
,
\nonumber \\
   \VEC{B}(\VEC{r},t>L/c)&=
    \sum {\rm Im} \bigl\{ \hat{A} \hat{\VEC{B}}(\VEC{r}) \exp(\ri\,\hat{\omega} t) \bigr\}
   +\VEC{B}_{\rm r}(\VEC{r},t)
,
\nonumber
\end{align*}
can be split into eigenfields and a residual part, $\VEC{E}_{\rm r}$ or $\VEC{B}_{\rm r}$. The long-range interaction between bunches or particles is essentially driven by the modal
part, as the residual fields decay or are not stimulated resonantly.
The  complex mode amplitudes are proportional to the source charge and depend on the source
offset. Hence they can be expressed as $\hat A=q_1 \hat{f}(x_1,y_1)$. 

Suppose that  a small test charge $\delta q$ follows the source particle on the same trajectory  at a distance
of $s>0$. It induces the additional amplitude
$\delta \hat{A}=\delta q \exp(-\ri\, \hat{\omega} s/c)   \hat{f}(x_1,y_1)$.
Therefore the energy of the modes is increased by
\begin{align*}
\delta {\cal W}_{\rm modes}&= \sum \bigl( |{\hat A}+\delta{\hat A}|^2 -|{\hat A}|^2   \bigr) \hat{\cal W}
\nonumber \\
                  &\approx \sum 2\,{\rm Re} \bigl\{ {\hat A} \delta{\hat A}^* \bigr\} \hat{\cal W}
\nonumber \\
                  &\approx 2 q_1 \delta q 
                 \sum  |\hat{f}(x_1,y_1)|^2 \,{\rm Re} \bigl\{   \exp(\ri\, \hat{\omega} s/c)  \bigr\}  \hat{\cal W}
                 .
\nonumber
\end{align*}
On the other hand, the test particle gains kinetic energy
\begin{align*}
\delta{\cal W}_{\rm k}&=\int_{-\infty}^{\infty} \delta q \,E_z(x_1,y_1,z-s,z/c) \,\rd z
\nonumber \\
          &=q_1 \delta q \sum {\rm Re} \left\{ \hat{f}(x_1,y_1) 
          \int_{-\infty}^{\infty} \hat{E}_{z}(x_1,y_1,z)\exp(\ri \, \hat \omega (z+s)/c) \,\rd z \right\}
         + \cdots         .
\nonumber
\end{align*}
The sum of the field energy and the kinetic energy is conserved, if terms with the same oscillation frequency  $\exp (\ri \, \hat \omega s/c)$  cancel:
$$
2|\hat{f}(x_1,y_1)|^2  \, \hat{\cal W}
+
\hat{f}(x_1,y_1)   \int_{-\infty}^{\infty} \hat{E}_{z}(x_1,y_1,z) \exp(\ri\, \hat \omega (z)/c) \,\rd z
=0.
$$
This is satisfied with $\hat{f}(x_1,y_1)=-\hat{v}^*(x_1,y_1)/\sqrt{\hat{\cal W}}$  and the normalized mode voltages
$$
\hat{v}(x,y)=\frac{1}{2\sqrt{\hat{\cal W}}}
\int_{-\infty}^{\infty} \hat{E}_z(x,y,z) \exp(\ri\, \hat \omega z/c) \,\rd z
,
$$
which do not depend on the arbitrary normalization mode fields.

The amplitude excited by the charge $q_1$ is
$$
\hat{A}=q_1 \hat{f}(x_1,y_1) =-q_1 \hat{v}^*(x_1,y_1)\big/\sqrt{\hat{\cal W}}
\: ,
$$
and the energy of all modes is 
$$
{\cal W}_{\rm EM,modes}=\sum |\hat A|^2 \hat W= q_1^2 \sum \hat k,
$$ with the per-mode loss parameter
$$
\hat{k}=|\hat{v}(x_1,y_1)|^2=\frac{1}{4 \hat{\cal W}} 
 \left| \int_{-\infty}^{\infty} \hat{E}_{z}(x_1,y_1,z) \exp(\ri\, \hat \omega (z)/c) \,\rd z \right|^2.
$$

The excitation of mode amplitudes depends linearly on the source distribution:
another particle with charge $q_2$ and offset $(x_2, y_2)$ at a  distance $s$ gives rise to
the additional amplitude
$$
\hat{A}=-q_2 \hat{v}^*(x_2,y_2) \exp (-\ri\, \hat{\omega} s/c) \big/\sqrt{\hat{\cal W}}
\: ,
$$
with phase shift $-\hat{\omega} s/c$ due to the time shift $s/c$. Therefore
it is possible to calculate the mode excitation for arbitrary charge distributions;
for example, for a one-dimensional bunch with offset $(x_1, y_1)$ and line charge
density $\lambda(z,t)=\lambda(z-ct)$, 
$$
\hat{A}=\hat{v}^*(x_1,y_1) \int \lambda (-s) \exp (-\ri\, \hat{\omega} s/c) \,\rd s
.
$$
In particular, a Gaussian bunch with charge $q_1$ and rms length $\sigma$ excites the amplitudes
$\hat{A}=q_1 \hat{v}^*(x_1,y_1) \exp (-(\hat{\omega} \sigma/c)^2/2)$. We introduce the 
shape-dependent per-mode loss parameter 
$$
\hat{k}_\sigma=\hat{k}\exp (-(\hat{\omega} \sigma/c)^2) .
$$
The electromagnetic field energy of all modes, after such a bunch has traversed the cavity,
is
$$
{\cal W}_{{\rm EM,modes,}\sigma}=\sum |\hat A|^2 \hat{\cal W}= q_1^2 \sum {\hat k}_\sigma .
$$

\subsection{Contribution of eigenmodes to the wake function}
After the source particle has traversed the cavity, the electromagnetic fields are
\begin{align*}
   \VEC{E}(\VEC{r},t>L/c)&= q_1
    \sum {\rm Re} \left\{ -\hat{v}^*(x_1,y_1) \hat{\cal W}^{-1/2} \hat{\VEC{E}}(\VEC{r})  \exp(\ri\, \hat{\omega} t) \right\}
   +\VEC{E}_{\rm r}(\VEC{r},t)
,
\nonumber \\
   \VEC{B}(\VEC{r},t>L/c)&= q_1
    \sum {\rm Im} \left\{ -\hat{v}^*(x_1,y_1) \hat{\cal W}^{-1/2}  \hat{\VEC{B}}(\VEC{r}) \exp(\ri\,\hat{\omega} t) \right\}
   +\VEC{B}_{\rm r}(\VEC{r},t)
,
\nonumber
\end{align*}
Therefore the momentum of a test charge $q_2$ at a distance $s>L$ behind the source,
with offset $(x_2, y_2)$ and velocity $c$, is changed by
$$
\Delta\VEC{p}=\frac{q_1 q_2}{c} \sum {\rm Re} \left\{- \hat{v}^*(x_1,y_1) \hat{\cal W}^{-1/2} \int_{-\infty}^{\infty} \rd z
   \left[\bigl( \hat{\VEC{E}} - \ri\,  \VEC{c} \times \hat{\VEC{B}} \bigr)
         \exp(\ri\,\hat{\omega} (z+s)/c)
   \right]
                               \right\}
+\Delta\VEC{p}_{\rm r}
,
$$
where the term $\Delta\VEC{p}_{\rm r}$ stands for the contribution of the residual fields. Likewise,
we can split the wake into modal and residual parts:
$$
\VEC{w}(x_1,y_1,x_2,y_2,s> L)= \frac{c\Delta\VEC{p}}{q_1 q_2}
              =\sum \hat{\VEC{w}}(x_1,y_1,x_2,y_2,s) + \VEC{w}_{\rm r}(x_1,y_1,x_2,y_2,s),
$$
where 
$$
\hat{\VEC{w}}(x_1,y_1,x_2,y_2,s>L)=-2 \,{\rm Re} \bigl\{ \hat{v}^*(x_1,y_1) \hat{\VEC{v}}(x_2,y_2)
                                               \exp(\ri\, \hat{\omega}s/c)
                                       \bigr\},
$$
with the normalized vectorial voltages
$$
   \hat{\VEC{v}}(x,y)=\frac{1}{2 \sqrt{\hat{\cal W}}}
   \int_{-\infty}^{\infty} \rd z
   \left[\bigl( \hat{\VEC{E}}(x,y,z) - \ri\,  \VEC{c} \times \hat{\VEC{B}}(x,y,z) \bigr)
         \exp(\ri\, \hat{\omega}z/c)
   \right]
.
$$

\subsection{Loss parameters}
Loss parameters describe the loss of energy of a source particle or source distribution
to electromagnetic field energy.

We have seen that the total energy loss of a point particle
is given by the wake function for $x_1=x_2$, $y_1=y_2$ and $s=0$, so that
$$
k_{\rm tot} = -w(x_1,y_1,x_2,y_2,0) = {\cal W}_{\rm EM,total}/q_1,
$$
and we know that the loss to eigenmodes is given by the per-mode loss parameters
$$
{\hat k}=|v(x_1,y_1)|^2 .
$$
The sum of all the per-mode loss parameters converges for cavities \emph{with} beam pipes
to a value below the total loss parameter $k_{\rm tot}$, as  not only are there modes
excited, but also field energy is scattered and propagates along the beam pipes.
(The wake of a closed cavity is completely determined by oscillating modes, but
the sum is divergent.)

The wake potential (of distributed sources) and the shape-dependent loss parameter
are usually calculated directly using electromagnetic time-domain solvers. 
The shape-dependent total loss parameter is the convolution of the longitudinal
wake potential with the charge density function; for instance, for bunches with
 longitudinal profile $\lambda(z,t)=\lambda(z-ct)$ and negligible
transverse dimensions,
$$
k_{{\rm tot},\sigma}= -\int_{-\infty}^{\infty} W(x_1,y_1,x_1,y_1,z) \lambda(z) \,\rd z.
$$
The excitation of eigenmodes by distributed sources was discussed in 
Section~\ref{section_Excitation_of_Eigenmodes}; the shape-dependent
per-mode loss parameter for a thin Gaussian bunch is 
$$
\hat{k}_\sigma=\hat{k} \exp \bigl( -(\hat{\omega}\sigma/c)^2 \bigr),
$$
and for a general longitudinal profile $\lambda$ it is
$$
\hat{k}_\sigma=\hat{k} \left| \int_{-\infty}^\infty \lambda(z) \cos (\hat{\omega} z/c) \,\rd z \right|^2 .
$$

For closed cavities, the sum of the per-mode loss parameters $\hat{k}_\sigma$ converges
to the total loss parameter $k_{{\rm tot},\sigma}$. This is also true for long bunches with
$\sigma \gg c/\omega_{\rm cutoff}$, which cannot excite frequencies above the lowest cutoff
frequency $\omega_{\rm cutoff} \propto \pi c/a$ of the beam pipes, where $a$ is the characteristic
transverse dimension of the pipes.

For the extreme case of ultra-short bunches it is difficult to calculate the wake potential, as
a very high spatial resolution is required. In this case, only a small fraction of energy is lost
to resonant modes and only a small part of the wakes is caused by resonances.

\section{Impedances}
\subsection{Definitions}
The Fourier transform of the negative\footnote{The sign is chosen so as to
obtain a non-negative real part for $x_1=x_2$ and $y_1=y_2$.}
wake function is called the impedance
or coupling-impedance:
$$
 Z_{\|}(x_1,y_1,x_2,y_2,\omega) = -\frac{1}{c}
\int_{-\infty}^{\infty}\rd s \, w_{\|}(x_1,y_1,x_2,y_2,s)
 \exp ( -\ri\, \omega s/c).
$$

\begin{figure}[hbtp]
  \begin{center}
  \resizebox{12cm}{!}{\includegraphics*{%
        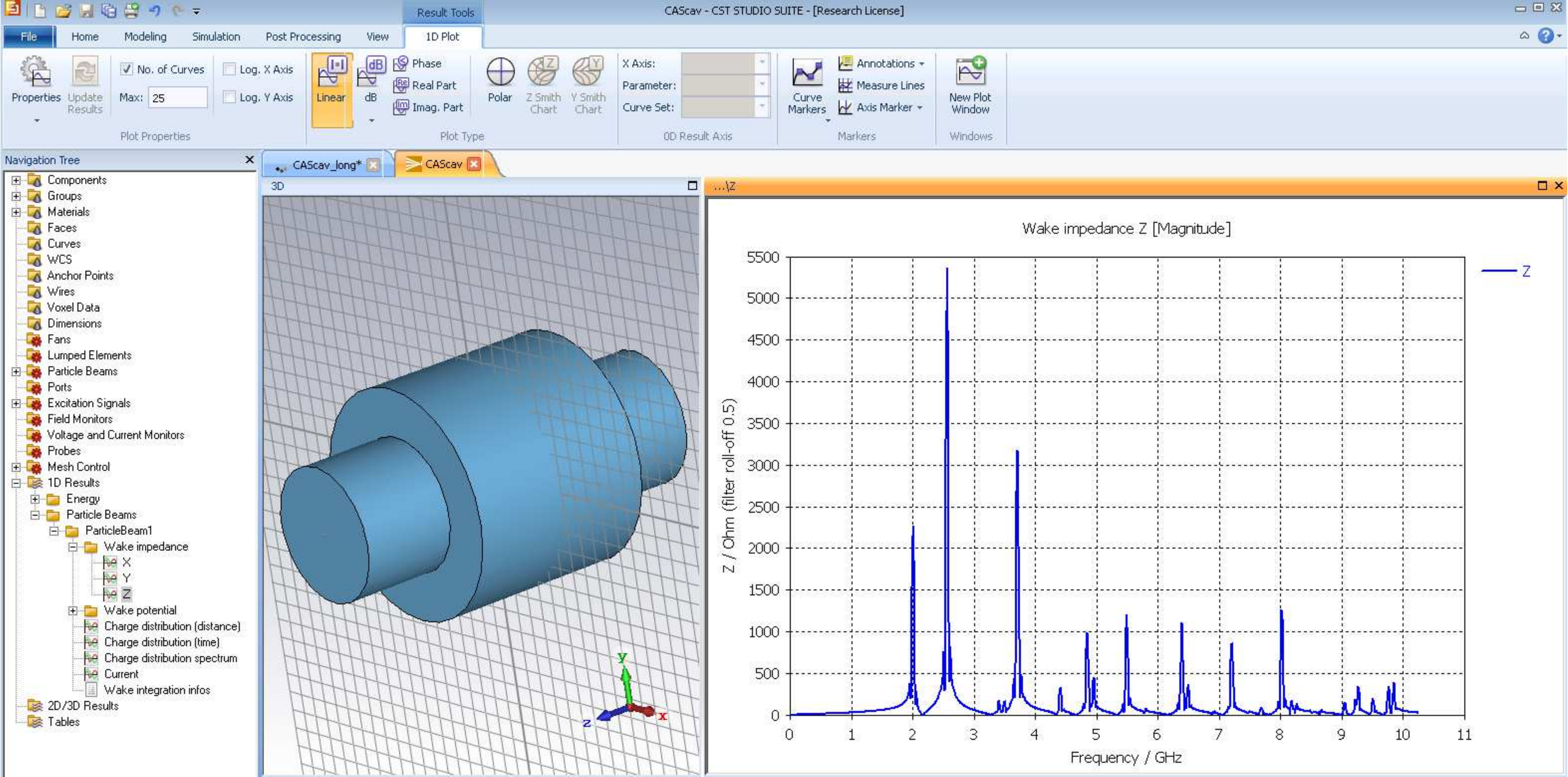}}
  \end{center}
\caption{
Real part of the impedance for a cavity with side pipes; 
the peaks correspond to cavity modes. The results were obtained with
the CST wakefield solver.
 }
\label{fig11}
\end{figure}

The wake function and impedance are two descriptions of the same thing, namely the
coupling between the beam and its environment. The wake function is the
time-domain description, while the impedance is the frequency-domain description:
$$
 w_{\|}(x_1,y_1,x_2,y_2,s) = -\frac{1}{2\pi}
\int_{-\infty}^{\infty}\rd\omega \, Z_{\|}(x_1,y_1,x_2,y_2,\omega)
\exp ( \ri\, \omega s/c).
$$
The reason for the usefulness of the impedance is that it often contains
a number of sharply defined frequencies corresponding to the modes of the
cavity or the long-range part of the wake. Figure~\ref{fig11}
shows the real part of the impedance for a cavity.
Below the cutoff frequency of the beam pipe, there is a sharp peak for
each cavity mode. The spectrum above the cutoff frequency is continuous,
caused by residual fields (not related to eigenmodes) and
by the `turn-on' of the harmonic eigen-oscillations. The
continuous part of the spectrum is important for short-range wakes,
especially for very short bunches.
\setlength{\unitlength}{1cm}

For the transverse impedance, it is often convenient to use a definition
containing an extra factor $ \ri $:
$$
 \VEC{Z_{\bot}}(x_1,y_1,x_2,y_2,\omega) = \frac{\ri}{c}
\int_{-\infty}^{\infty}\rd s \, \VEC{w_{\bot}}(x_1,y_1,x_2,y_2,s) 
      \exp ( -\ri\, \omega s/c).
$$
The reason is that the transverse--longitudinal relations due to the
Panofsky--Wenzel theorem then read as follows in the frequency domain:
$$
\frac{\omega}{c} \,  \VEC{Z_{\bot}}(x_1,y_1,x_2,y_2,\omega) =
 \left( \VEC{e}_x \frac{\partial}{\partial x_2}+\VEC{e}_y \frac{\partial}{\partial y_2} \right) 
Z_{\|}(x_1,y_1,x_2,y_2,\omega).
$$

\subsection{Some properties of impedances and wakes}
In the spatial $s$-domain, the relationship between the wake potential of a line
charge density $\lambda(z-ct)$ and the wake functions of a point particle
is described by the convolution
$$
W(x_1,y_1,x_2,y_2,s)=\int_{- \infty}^\infty \rd z \: w(x_1,y_1,x_2,y_2,s+z) \lambda(z).
$$
The corresponding equation in the frequency domain for the Fourier
transform of the negative wake potential  is
$V(x_1,y_1,x_2,y_2,\omega)=Z_\|(x_1,y_1,x_2,y_2,\omega) I(\omega)$,
where 
$$
I(\omega)=\int_{-\infty}^{\infty} i(t) \exp(-\ri\,\omega t) \,\rd t
         =\int_{-\infty}^{\infty} c \lambda(-ct) \exp(-\ri\,\omega t) \,\rd t
$$
is the beam current in the frequency domain. The energy loss of the bunch to electromagnetic fields,
\begin{align*}
{\cal W}_{\rm loss}&=\int_{-\infty}^\infty W(x_1,y_1,x_1,y_1,s) \lambda(-s) \,\rd s
\nonumber \\
            &= \frac{1}{2\pi} \int_{-\infty}^\infty V(x_1,y_1,x_1,y_1,\omega) I(\omega)^* \,\rd\omega
\nonumber \\
            &= \frac{1}{\pi} \int_{0}^\infty {\rm Re}\{Z(x_1,y_1,x_1,y_1,\omega) \} |I(\omega)|^2 \,\rd\omega,
\nonumber
\end{align*}
has to be non-negative for any bunch shape $\lambda$. Therefore the real part of the longitudinal impedance must  be non-negative for all offsets with $x_1=x_2$ and $y_1=y_2$. The real part can be negative for, say, 
$x_1=-x_2$ and $y_1=-y_2$, for a structure with azimuthal symmetry and frequency close to a dipole
resonance.

As the wake potential is a real function,  the real part of the impedance is an even function of the
frequency while the imaginary part is an odd function of it:
$$
 {\rm Re} \{ Z_{\|}(\ldots ,\omega) \} =
 {\rm Re} \{ Z_{\|}(\ldots ,-\omega) \}, \qquad 
 {\rm Im} \{ Z_{\|}(\ldots ,\omega) \} = - 
 {\rm Im}\{ Z_{\|}(\ldots ,-\omega) \}.
$$
Hence, the wake function is given in terms of the impedance as 
\begin{align*}
w_{\|}(\ldots ,s) & =   -\frac{1}{2\, \pi} \int_{-\infty}^{\infty} \rd\omega \,
Z_{\|} (\ldots ,\omega) \exp ( \ri\, \omega s/c)
\nonumber \\
 & =   - \frac{1}{2\, \pi} \int_{-\infty}^{\infty} \rd \omega \,
\bigl( {\rm Re} \{ Z_{\|}(\ldots ,\omega) \} 
                \cos(\omega s/c)- 
{\rm Im} \{ Z_{\|}(\ldots ,\omega)\}
                \sin(\omega s/c) 
\bigr).
\end{align*}
Furthermore, the electromagnetic field ahead of the source particle is zero for $v=c$,
as electromagnetic waves cannot overtake the source. Therefore, the wake function is
\emph{causal}, and the real and imaginary parts of the impedance are dependent on each other.
From $ w_{\|}(\ldots ,s<0) = 0 $ it follows that  for $u=-s>0$,
$$
 \int_{-\infty}^{\infty} \rd \omega \,
{\rm Re} \{ Z_{\|}(\ldots ,\omega) \} 
        \cos(\omega u/c)
=
 - \int_{-\infty}^{\infty} \rd \omega \,
 {\rm Im} \{ Z_{\|}(\ldots ,\omega) \} 
        \sin(\omega u/c),
$$
so only the real (or imaginary) part of the impedance is really needed:
$$
w_{\|}(\ldots ,s>0) =  \frac{1}{ \pi} \int_{-\infty}^{\infty} \rd \omega \,
 {\rm Re} \{ Z_{\|}(\ldots ,\omega) \} 
                \cos(\omega s/c) .
$$

%
%

%
\subsection{Shunt impedance and quality factor}
The modal part of the wake function is
$$
\hat{\VEC{w}}(x_1,y_1,x_2,y_2,s)=-2 \,{\rm Re} \bigl\{ \hat{v}^*(x_1,y_1) \hat{\VEC{v}}(x_2,y_2)
                                               \exp(\ri\, \hat{\omega}s/c)
                                       \bigr\}
                             \begin{cases}
                                           0 & \mbox{ for } s<0, \\
                                           1 & \mbox{ for } s=0, \\
                                           2 & \mbox{ otherwise}.
                                     \end{cases}
$$
We are interested in the longitudinal component on the axis ($x_1=y_1=x_2=y_2=0$), and 
for simplicity we omit the transverse coordinates, so
$$
\hat{w}_\|(s)=-2 \hat{k} \cos ( \hat{\omega} s/c)
                              \begin{cases}
                                           0 & \mbox{ for } s<0, \\
                                           1 & \mbox{ for } s=0, \\
                                           2 & \mbox{ otherwise}.
                                     \end{cases}
$$
Then the impedance per mode,
$$
\hat{Z}_\|(\omega)=-{1 \over c} \int_{-\infty}^\infty \hat{w}_\|(s) \exp(-\ri\, \omega s/c) \,\rd s,
$$
is calculated as
$$
\hat{Z}_\|(\omega)=2\hat{k} \left\{ \pi \delta(\omega+\hat{\omega})+\pi \delta(\omega-\hat{\omega})
                             + {\ri\, \omega \over \hat{\omega}^2-\omega^2}
                      \right\}.
$$
This is equivalent to the impedance of a parallel resonant circuit (see Fig.~\ref{fig12}),
$$
\hat{Z}_\|(\omega)= \lim_{\hat{R} \to \infty } 
      \left(\ri\, \omega \hat{C} + \frac{1}{\ri\, \omega \hat{L}} + \frac{1}{\hat{R}} \right)^{-1}
$$
with $\hat{C}=1/(2\hat{k})$, $\hat{L}=2\hat{k}/\hat{\omega}^2$ and $\hat{R} \to \infty$.
\setlength{\unitlength}{1cm}
\begin{figure}[hbtp]
  \begin{center}
    \begin{picture}(4,3)(0,0)                                                            
        \put(0,0){\resizebox{4cm}{!}{\includegraphics*{%
       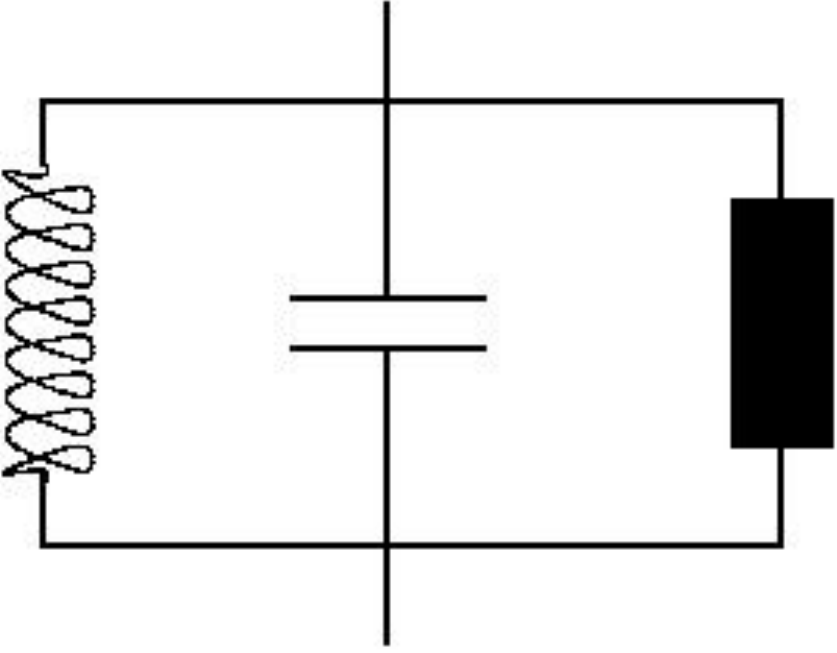}} }                      
      \put(0.5,2){\makebox(0,0)[bl]{ $L$}} 
      \put(2,2){\makebox(0,0)[bl]{ $C$}} 
      \put(4,2){\makebox(0,0)[bl]{ $R$}}                                             
    \end{picture}     
  \end{center}
\caption{
Equivalent circuit model of the impedance of one mode
 }
\label{fig12}
\end{figure}
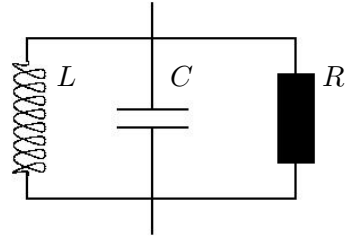

Although the resistor $\hat{R}$
was introduced for obvious formal reasons, it is helpful to consider weak losses of a
resonator with a high quality factor $\hat{Q}=\hat{R}/(\hat{\omega}\hat{L})$. The impedance per
mode of a resonator with weak losses is
$$
\hat{Z}_\|(\omega)=2 \hat{k}\, \frac{\ri\, \omega}{\hat{\omega}^2-\omega^2+\ri\,\omega\hat{\omega}/\hat{Q}}.
$$
The resistor $R=\hat{Z}_\|(\hat{\omega})=2 \hat{k} \hat{Q} / \hat{\omega}$ is called
the \emph{shunt impedance}. The \emph{quality factor} $\hat{Q}$ describes the decay time
$$
\hat{\tau}=2\hat{Q}/\hat{\omega} ,
$$
the resonance bandwidth
$$
\Delta \hat{\omega}=\omega/\hat{Q},
$$
with $|\hat{Z}_\|(\hat{\omega})/\hat{Z}_\|(\hat{\omega}\pm \Delta \hat{\omega}/2)|^2\approx 2$, 
and the energy loss per unit time
$$
\hat{P}=\hat{\omega} \hat{\cal W}/\hat{Q}.
$$
The last relation is used to determine the quality factor by perturbation theory: the energy loss
(without beam) is caused by wall losses; as a good approximation these can be calculated from
the fields obtained for the mode with infinite conductivity. If $\hat{H}_{\rm t}$ is the
magnetic field tangential to the surface, the total power dissipated into the wall is
given by a surface integral
$$
\hat{P}={1 \over 2} \int_{\partial V} {\rm Re} \bigl\{ \hat{\VEC{E}}_{\rm s} \times \hat{\VEC{H}}^* \bigr\} \cdot \rd\VEC{A}
       ={1 \over 2} \int_{\partial V}  \sqrt{\frac{\hat{\omega} \mu }{2 \kappa}}\: | \hat{\VEC{H}} |^2 \,\rd A
$$
where $\hat{\VEC{E}}_{\rm s}=Z_{\rm s} \VEC{n} \times \hat{H}$ is  the tangential component of the electric
field on the surface, with  surface impedance $Z_{\rm s}=\sqrt{\ri\,\hat{\omega}\mu/\kappa}$ for conductivity~$\kappa$.

\section{Instabilities}
\subsection{Equation of motion}
In the previous section the properties of the wake potential
and the related impedance has been discussed in detail.
Now the effect on the beam motion will be studied.
The kick on a test charge due to a transverse dipole wake is
$$
\Vec{\theta}(s) = \frac{e}{E} \, q \,   W^{(1)}_{\bot}(s) \, \Vec{r},
$$
where $E$ is the beam energy, $q$ the charge of the bunch with
transverse offset  $\Vec{r}$ .

In the rigid bunch approximation the transverse equation of a bunch
in a storage can be written as:
$$
\frac{d^2}{ds^2} y(s)
   + { \left( \frac{ \omega_{\beta} }{c} \right) }^2 \,\, y(s)
 =  0,
 \label{free-oscillation}
$$
were $s$ is the longitudinal position in the storage ring,
$y$ is a transverse coordinate of the bunch and
$ \omega_{\beta}$ the betatron frequency.

In order to include wakefield effects one has to modify the
above equation \cite{Chao93}:
$$
 \frac{d^2}{ds^2} y(s)
   + { \left( \frac{ \omega_{\beta} }{c} \right) }^2 \,\, y(s)
 =  \frac{e\,\,\, q}{m c^2 \, \gamma} 
     \sum^{\infty}_{n=0}
     \frac{1}{C} \,\, y(s - n \,C ) \,\,\,\,
     W^{(1)}_{\bot}(n \,C ),
$$
where $C$ is the circumference of the storage ring and $\gamma$
the relativistic $\gamma$-factor. The right hand side of the
equation is a sum of all the wakefield kicks during each turn in the
storage ring.

An ansatz for the solution of the above equation is 
$$
y(s) \sim \, \exp(-i \, \Omega \, s/c),
$$
with the complex frequency $ \Omega$:
$$
 \Omega = \omega_{\beta} + i \, \frac{1}{\tau},
$$
including the betatron frequency $\omega_{\beta}$
and the growth or damping rate $1/\tau$.

The ansatz leads to a relation for the complex frequency $ \Omega$:
$$
\Omega^2 - \omega_{\beta}^2 = 
       -i \, \frac{e\,\,\, q}{m c^2 \, \gamma}  \,
         \frac{c}{T_0^2} \,
               \sum^{\infty}_{p=-\infty}
         Z^{(1)}_{\bot} (\Omega + p \omega_0).
$$         

The transformation from the time domain picture to the
frequency picture is based on the relation:
$$
 \sum^{\infty}_{n=0} \exp(i \,\, n \,\, \Omega \, T_0) \,\,
      W^{(1)}_{\bot}(n \,C )
    = \frac{i}{T_0} \, 
      \sum^{\infty}_{p=-\infty}
      Z^{(1)}_{\bot} (\Omega + p \omega_0),
$$
which allows to replace the wake potential with the
impedance $ Z^{(1)}_{\bot}$. $T_0$ is the revolution time in the
storage ring.

If one further assumes that $\Omega$ does not deviate much from $ \omega_{\beta}$
($ \Omega+\omega_{\beta} \approx 2 \, \omega_{\beta}$) one
obtains an relation for the betatron tune shift and the growth rate with the
transverse impedance:
$$
\Omega - \omega_{\beta} = 
       -i \, \frac{1}{2 \omega_{\beta}} \, \frac{e\,\,\, q}{m c^2 \, \gamma}  \,
         \frac{c}{T_0^2} \,
               \sum^{\infty}_{p=-\infty}
         Z^{(1)}_{\bot} (\omega_{\beta} + p \omega_0).
$$
The imaginary part is contributing to a mode frequency shift
$\Delta \Omega = {\rm Re}(\Omega - \omega_{\beta}$, while the real part
of the impedance corresponds to a growth rate
$1/{\tau} = {\rm Im}(\Omega - \omega_{\beta})$:
$$
\tau^{-1} =  - \, \frac{1}{2 \omega_{\beta}} \, \frac{e\,\,\, q}{m c^2 \, \gamma}  \,
         \frac{c}{T_0^2} \,
               \sum^{\infty}_{p=-\infty}
         {\rm Re}(Z^{(1)}_{\bot}) (\omega_{\beta} + p \omega_0).
$$         
This demonstrates how the impedance can be use to calculate instability growth rates
 for a simplified model of transverse rigid bunch motion.
 In the next section a two macro particle is used to get a basic understanding
 of a head tail instability in an storage ring.

\subsection{Head tail instability}
\begin{figure}[hhhhhtbp]
   \begin{center}
    \rotatebox{0}{\resizebox{0.46\textwidth}{!}{%
          \includegraphics*{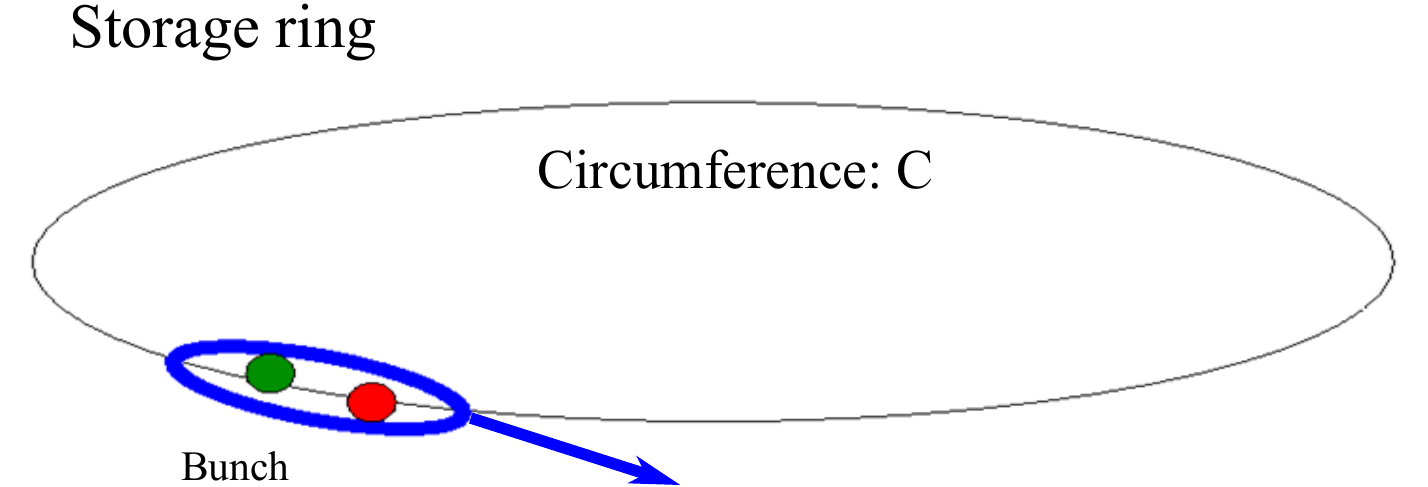}}}
   \end{center}
\caption{
Headtail model of a bunch in a storage ring.
 }
\label{fig_headtail1}
\end{figure}

In the previous section the bunch was modelled
just as a rigid bunch without an internal structure.
Now it is assumed that the bunch consists
of two parts, a head and a tail as show in Fig.~\ref{fig_headtail1}.
This two particle model can be used to included synchrotron
oscillations.
\setlength{\unitlength}{1cm}

The starting point is again the kick of the head particle
on the tail particle of the bunch:
The kick on a test charge due to a transverse dipole wake is
$$
\theta_{tail} = \frac{e}{E} \, \frac{q}{2} \,   {\cal W}_{\bot} \, y_{head}
$$
where $E$ is the beam energy, $q$ the charge of the bunch with
transverse offset  $y_{head}$ of the head. $ {\cal W}_{\bot}$ is the
effective wake of the head of the bunch acting on the tail of the bunch.

\begin{figure}[thbp]
   \begin{center}
    \rotatebox{0}{\resizebox{0.6\textwidth}{!}{%
          \includegraphics*{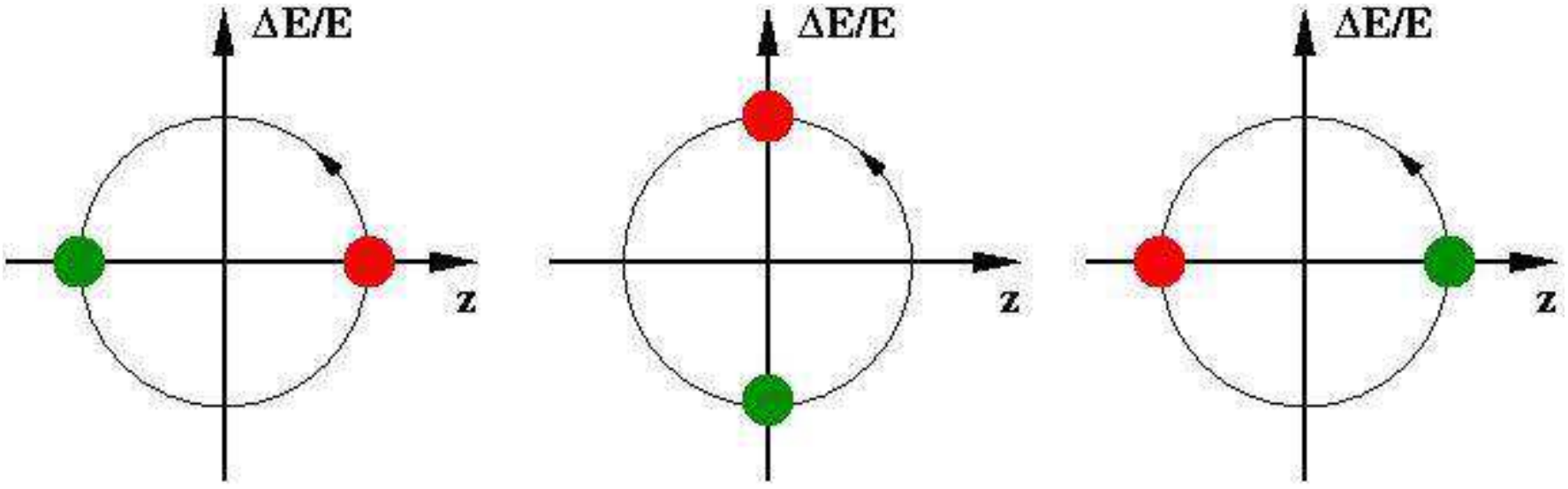}}}
   \end{center}
\caption{
Synchrotron oscillation of the head and tail macro particles in the bunch.
 }
\label{fig_headtail2}
\end{figure}

The equation of motion for the two particles are:
$$
{y''}_1 + \left(\frac{ \omega_{\beta}}{c}\right)^2 y_1 = 0
$$
for the head particle and
$$
{y''}_2 + \left(\frac{ \omega_{\beta}}{c}\right)^2 y_2 = 
\frac{N \, r_0}{2 \gamma \, C} \, {\cal W}_{\bot} \, y_1
$$
for the tail particle, where 
$$
r_0 = \frac{1}{4 \pi \, \epsilon_0} \, \frac{e^2}{m_0\, c^2} = 2.818 \, \cdot \, 10^{-15} \, {\rm m}
$$
is the classical radius of the electron and $N$ is the bunch population.
$N/2$ electrons are in the head.
The positions of the head and the tail are exchanged after one synchrotron oscillation
period as illustrated in Fig.~\ref{fig_headtail2}. Therefor the above equation of motion
are only valid for one half of a period of synchrotron oscillations. For the second half of the
synchrotron oscillation period the indices's have to interchanged.

\setlength{\unitlength}{1cm}

The equation of motions can also be rewritten in a complex phasor notion
for both particles with the index 1 and 2 \cite{Chao93}:
$$
\tilde{y}_{1,2} = y_{1,2} + i \, \frac{c}{\omega_{\beta}} {y'}_{1,2}
$$
For the first half of the synchrotron oscillation period the following relations holds:
$$
{\left( \begin{array}{c} \tilde{y}_1\\
                        \tilde{y}_2
       \end{array}   \right)}_{s= c \, T_s/2} =
\exp( -i \ \omega_{\beta} \, T_s/2)
\left( \begin{array}{cc}     1 &  0  \\
                          i \, \Upsilon &  1
       \end{array}   \right) \,
{\left( \begin{array}{c} \tilde{y}_1\\
                         \tilde{y}_2
       \end{array}   \right)}_{s=0}, 
$$       
where the wakefield effect is now presented with the parameter $\Upsilon$,
defined as:
$$
\Upsilon = \frac{\pi \, N \, r_0 \, c^2}{4 \gamma \, C \omega_{\beta} \omega_s} \, {\cal W}_{\bot} .
$$

For one complete synchrotron period one obtains now the relation
$$
{\left( \begin{array}{c} \tilde{y}_1\\
                        \tilde{y}_2
       \end{array}   \right)}_{s= c \, T_s} =
\exp( -i \ \omega_{\beta} \, T_s)\,
\left( \begin{array}{cc}     1 &  i \, \Upsilon   \\
                             0 &  1
       \end{array}   \right) \,
\left( \begin{array}{cc}     1 &  0  \\
                          i \, \Upsilon &  1
       \end{array}   \right) \,
{\left( \begin{array}{c} \tilde{y}_1\\
                         \tilde{y}_2
       \end{array}   \right)}_{s=0}.
$$
Stability requires pure imaginary eigenvalues of the product matrix which can be translated
into a criteria for the parameter $\Upsilon$:
$$
\Upsilon < 2.
$$
For a known effective wakefield  ${\cal W}_{\bot}$ this can be directly translated into an
limit for the bunch population which will be stable with respect to head tail instabilities:
$$
N <  \frac{2 \, \gamma \, C \omega_{\beta} \omega_s}{\pi \, N \, r_0 \, c^2} \, \frac{1}{{\cal W}_{\bot}}.
$$

\subsection{Ion trapping}
Finally, the effect of ion trapping will be discussed, which can
result in an increased beam emittance, betatron tune shifts
and reduced beam lifetime.
The effect of the ion cloud on the beam can be
modelled as a broad band resonator wake field \cite{Wang2011}.
One can apply the classical theory of instabilities to obtain 
stability criteria with respect to ion effects.
Nevertheless, it is interesting to have a simple criteria at hand to
know whether one can expect trapped ions in a beam.
This was already analyzed by Kohaupt in 1971 \cite{Kohaupt1971}
for the storage ring DORIS at DESY.

The residual gas density $d_{gas}$  at room temperature (300 K) can be calculated
from vacuum pressure:
$$
d_{gas} = \frac{p_{gas} \,\, N_{Avo}}{R_{gas} \,\, 300 \,{\rm K} } = 24.14 \, \cdot \, 10^6 \, {\rm cm}^{-3},
$$
where the numbers have been calculated for $p_{gas} = 1 \, \cdot \, 10^{-9} \,\, {\rm mbar}$.
$R_{gas} = 8.31447\, {\rm J/(K \, mol)}$ is the general gas constant and $N_A = 6.0221367 \cdot 10^{23}$
is the Avogadro number.

Assuming now a typical cross section of 2~Mbarn ($2  \cdot 10^{-18}\, {\rm cm}^2$) for the ionization
process, one obtains an ion density of
$$
\lambda_{ion} = d_{gas} \, \sigma_{ion} \, N_0 \approx 2 \, {\rm Mbarn} \,  d_{gas} \, N_0,
$$
where $N_0$ is the bunch population. After the passage of one bunch with a bunch population
of $5 \cdot 10^{9}$ the ion density is already $0.24$~ions/cm or $230$~ions/cm after the
passage of 960 bunches, which is one of the filling modes of the synchrotron radiation facility
PETRA~III at DESY.

The bunch train in a storage ring is acting as sequence of quadrupole lenses on the ion beam.
In a linear approximation the interaction of the bunch with the ion can be can be presented by a
matrix:
$$
M = \left( \begin{array}{cc}  1 &  L_b  \\
                          0 &  1
       \end{array}   \right) \,
\left( \begin{array}{cc}     1 &  0  \\
                          -a &  1
       \end{array}   \right) ,
$$
 where $L_b = c \, \Delta  t $ is the bunch spacing, and 
$$
 a = N_b \frac{2 \, r_p}{\sigma_y \, (\sigma_x + \sigma_y )} \, \frac{1}{A}
$$
is the linear force of the beam on the ion.
$N_b$ is the bunch population, $r_p=1.535 \cdot 10^{-18}$~m,
$\sigma_x, \,\sigma_y$ are the transverse rms beam dimensions and $A$ is
the mass number of the ion. For ${\rm CO}$ and ${\rm N}_2$ the mass number
is $A=28$, while for water $A=18$.
The ion motion will be only stable if the trace of the matrix $M$ is smaller than 2, or
$ Tr(M) = 2 - a\, L_b <2$. The ion will be trapped, when the mass number A
is larger than a critical mass number $A_c$:
$$
A > A_c = N_b \, L_b  \,  \frac{r_p}{2\, \sigma_y \, (\sigma_x + \sigma_y )}.
$$

\clearpage
\newpage

\addcontentsline{toc}{section}{\protect\numberline{}{Appendix A} }
\section*{Appendix A: Eigenmodes of a closed cavity}

We consider a (simply connected) cavity volume $V_{\rm c}$, with perfectly conducting walls
(boundary $\partial V_{\rm c}$) and without current density. We search for
time-harmonic eigensolutions, which can be written as
\begin{align*}
   \VEC{E}(\VEC{r},t)&=\hat{\VEC{E}}(\VEC{r}) \cos( \hat{\omega} t)
,
\nonumber \\
   \VEC{B}(\VEC{r},t)&=\hat{\VEC{B}}(\VEC{r}) \sin( \hat{\omega} t)
,
\nonumber
\end{align*}
where $\hat{\VEC{E}}$ and $\hat{\VEC{B}}$ are the eigenfields and $\hat{\omega}$ the (angular) eigenfrequencies.
Substituting these into Maxwell's equations gives
\begin{align*}
   \nabla \varepsilon \hat{\VEC{E}} &= \hat{\rho}
,
\nonumber \\
   \nabla \times \hat{\VEC{E}} &= -\hat{\omega} \hat{\VEC{B}}
,
\nonumber \\
   \nabla \hat{\VEC{B}} &= 0
,
\nonumber \\
   \nabla \times \mu^{-1} \hat{\VEC{B}} &= -\hat{\omega} \varepsilon \hat{\VEC{E}}
.
\nonumber
\end{align*}
We apply the operator $\nabla \times \mu^{-1}$ to the first curl equation and use the
second curl equation to eliminate the magnetic flux density, thus obtaining the eigenproblem
$$
\varepsilon^{-1}\nabla \times \mu^{-1} \nabla \times \hat{\VEC{E}} = \hat{\lambda}\hat{\VEC{E}}
,
$$
with the eigenvalues $\hat{\lambda}=\hat{\omega}^2$ and the boundary condition
$\VEC{n} \times \hat{\VEC{E}} = \VEC{0}$. The operator $\varepsilon^{-1} \nabla
\!\times\! \mu^{-1} \nabla \times$ is self-adjoint\footnote{
The property
$ \langle \varepsilon^{-1}\nabla \times \mu^{-1} \nabla \times\VEC{A},  \VEC{B} \rangle 
= \langle \VEC{A}, \varepsilon^{-1}\nabla \times \mu^{-1} \nabla \times \VEC{B} \rangle$
can be shown with help of the identity 
$ \nabla [ \VEC{A} \times \mu^{-1} \nabla \times \VEC{B} - \VEC{B} \times \mu^{-1} \nabla \times \VEC{A} ]
= \VEC{B} \times \nabla \times \mu^{-1} \nabla \times \VEC{A} - \VEC{A} \times \nabla \times \mu^{-1} \nabla \times \VEC{B}$
and the divergence theorem. The left-hand side gives a surface integral that is zero
because of  the boundary conditions. The right-hand side corresponds to the
assertion.
}
with scalar product
$$
\langle \VEC{A}, \VEC{B} \rangle = {1 \over 2} \int_{V_{\rm c}} \varepsilon \VEC{A} \cdot \VEC{B} \, \rd V
.
$$
Therefore the problem has an infinite number of discrete real eigenvalues and a complete
orthogonal system of eigenvectors,
$$
\langle \hat{\VEC{E}}_\xi, \hat{\VEC{E}_\tau} \rangle = \hat{\cal W}_\xi  \delta_{\xi\tau}
,
$$
where $\hat{\cal W}_\xi$ is the electromagnetic field energy of mode $\xi$. The eigenvalues
$\hat{\lambda}$ are non-negative so that all eigenfrequencies $\hat{\omega}$ are
real.\footnote{
This property can be shown by using the identity
$\nabla[\hat{\VEC{E}} \times \mu^{-1} \nabla \times \hat{\VEC{E}}]=
\mu^{-1}(\nabla \times \hat{\VEC{E}})^2
-\hat{\VEC{E}} \nabla \times \mu^{-1} \nabla \times \hat{\VEC{E}} $
and the divergence theorem. The left-hand side gives a surface integral that is zero
because of the boundary conditions. The volume integral of the first term on the
right-hand side is non-negative; the integral of the second term gives
$-2 \hat{\lambda} \hat{\cal W}$. As $\hat{\cal W}$ is positive, $\hat{\lambda}$ cannot be negative.
}

There are obviously two types of eigensolutions: 
\begin{alignat*}{4}
   \hat{\omega}&=0
   , &\qquad 
    \hat{\omega} &\ne 0, \\
   \nabla \varepsilon \hat{\VEC{E}} &\not\equiv 0
, &\qquad 
\nabla \varepsilon \hat{\VEC{E}} &= 0,\\
   \nabla \times \hat{\VEC{E}} &= 0
   , &\qquad 
  \nabla \times \hat{\VEC{E}} &= -\hat{\omega} \hat{\VEC{B}},\\
   \hat{\VEC{B}} &\equiv \VEC{0}
   , &\qquad 
  \hat{\VEC{B}} &\not\equiv \VEC{0}.
\end{alignat*}
Eigenfields for $\hat{\omega}=0$ are curl-free and are just solutions to the
electrostatic problem for any source distribution $\hat{\rho}$ and the boundary
condition $\VEC{n} \times \hat{\VEC{E}}=0$. Oscillating eigenfields are free of
divergence; this is a consequence of Maxwell's second curl equation.

In Appendix B we use the property that any linear combination of
eigensolutions with $\hat{\omega}=0$ is orthogonal to any linear combination
of oscillating eigenfields.

\addcontentsline{toc}{section}{\protect\numberline{}{Appendix B} }
\section*{Appendix B: Wake of a closed cavity}

We consider a (simply connected) cavity volume $V_{\rm c}$ of arbitrary shape,
with perfectly conducting walls, that is located between the planes $z=0$ and $z=L$.
It is traversed by a point particle with charge $q_1$, offset $(x_1, y_1)$ and  velocity $v=c$.
The stimulating charge and current density are
\begin{align*}
\rho(\VEC{r},t)&=q_1 \delta(x-x_1) \delta(y-y_1) \delta(z-ct)
, \nonumber \\
\VEC{j}(\VEC{r},t)&= c \VEC{e}_z \rho(\VEC{r},t)
. \nonumber
\end{align*}

We use the complete orthogonal system of eigensolutions to describe the time-dependent
electric field:
$$
\VEC{E}(\VEC{r},t)= \sum_{\nu \in C} \hat{a}_\nu(t) \hat{\VEC{E}}_\nu(\VEC{r})
,
$$
where $\nu$ is the mode index, $C$ is the set of all indexes and
 $\hat{a}_\nu(t)$ are the  time-dependent coefficients. 
 As in the main text, we shall write all
mode-specific quantities with a hat and omit the index $\nu$.
We solve Maxwell's equations
\begin{align*}
   \nabla \varepsilon \VEC{E} &=  \rho
,
\nonumber \\
   \nabla \times \VEC{E} &= - {\partial \over \partial t} \VEC{B}
,
\nonumber \\
   \nabla \VEC{B} &= 0
,
\nonumber \\
   \nabla \times \mu^{-1} \VEC{B} &= \VEC{J} + \varepsilon {\partial \over \partial t} \VEC{E}
\nonumber
\end{align*}
by applying the operator $\varepsilon^{-1} \nabla \times \mu^{-1}$ to the first curl equation
and eliminating the magnetic flux density with the help of the second curl equation:
$$
\varepsilon^{-1} \nabla \times  \mu^{-1} \nabla \times \VEC{E}= 
- \varepsilon^{-1} {\partial \over \partial t} \VEC{J} - {\partial^2 \over \partial t^2} \VEC{E}.
$$
By using the modal expansion and the eigenmode equation, we obtain 
$$
\sum_{\nu \in C} \hat{a}(t) \hat{\omega}^2 \hat{\VEC{E}}= 
- \varepsilon^{-1} {\partial \over \partial t} \VEC{J}
- {\partial^2 \over \partial t^2} \sum_{\nu \in C} \hat{a}(t) \hat{\VEC{E}} .
$$
This set of scalar equations can be decoupled by applying the operator
$\langle \hat{\VEC{E}}_\xi, \cdots  \rangle$ to both sides and using the orthogonality
condition:
$$
\hat{a}_\xi(t) \hat{\omega}_\xi^2 \hat{\cal W}_\xi= 
- \varepsilon^{-1} {\partial \over \partial t} \langle \VEC{\hat{E}}_\xi, \VEC{J} \rangle
- {\partial^2 \over \partial t^2}  \hat{a}_\xi(t) \hat{\cal W}_\xi.
$$
Finally, we substitute the Dirac current density and suppress the index, to arrive at
$$
\left( \hat{\omega}^2 + {\partial^2 \over \partial t^2} \right) \hat{a}(t) = 
{-1 \over \hat{\cal W} \varepsilon}\, {\partial \over \partial t} \langle \VEC{\hat{E}}, \VEC{J} \rangle
=-{c q_1 \over 2\hat{\cal W}} \, {\partial \over \partial t} \hat{E}(x_1,y_1,ct).
$$
This ordinary differential equation can be solved\footnote{
The causal solution of $\ddot{a}+\omega^2 a=\dot b$ is
$a(t)={\rm Re} \{ \int_{-\infty}^t b(\tau) \exp(\ri\,\omega (t-\tau))\,\rd\tau \}$.
} to give 
$$
\hat{a}(t)= {-q \over \sqrt{\hat{\cal W}}} \,{\rm Re} \bigl\{ \hat{v}^{*}(x_1,y_1,ct) \exp(\ri\, \hat{\omega} t) \bigr\}
$$
with
$$
v(x,y,z)=\frac{1}{2 \sqrt{\hat{\cal W}}} \int_{-\infty}^{z} \hat{E}_z(x,y,s) \exp(\ri\, \hat{\omega}s/c) \,\rd s
$$
and
$$
{\partial \over \partial z} v(x,y,z)=\frac{1}{2 \sqrt{\hat{\cal W}}}\,  \hat{E}_z(x,y,z) \exp(\ri\, \hat{\omega}z/c) .
$$

The longitudinal wake function is the sum over all modes,
$$
w_\|(x_1,y_1,x_2,y_2,s) = \sum_{\nu \in C} \hat{w}_\| (x_1,y_1,x_2,y_2,s),
$$
with the `per-mode' contributions
\begin{align*}
\hat{w}_\| (x_1,y_1,x_2,y_2,s) &={1 \over q_1} \int_{-\infty}^\infty \hat{a}\left(\frac{z+s}{c}\right) \hat{E}_z(x_2,y_2,z) \,\rd z
\nonumber \\                   &={-1 \over \sqrt{\hat{\cal W}}} 
\int_{-\infty}^\infty {\rm Re} \left\{  \hat{v}^*(x_1,y_1,z+s)
              \exp \left( \ri\, \hat{\omega} \frac{z+s}{c} \right) 
              \right\} \hat{E}_z(x_2,y_2,z) \,\rd z
\nonumber \\                   &=-2 \,
 {\rm Re} \left\{  \exp ( \ri\, \hat{\omega} s/c ) 
                \int_{-\infty}^\infty \hat{v}^*(x_1,y_1,z+s)
              {\partial \over \partial z} v(x_2,y_2,z) \,\rd z \right\}  .
\end{align*}
None of these terms is causal, i.e.\ $\hat{w}_\|(x_1,y_1,x_2,y_2,s<0)\not \equiv 0$, but the sum has to be!
In the following we use causality to find the simplified representation of the longitudinal wake
function 
$$
w_\|(x_1,y_1,x_1,y_1,s>0) = -2 \sum_{\hat{\omega}\ne0} \hat{k}(x_1,y_1) \cos (\hat{\omega} s/c)
$$
for $x_1=x_2$ and $y_1=y_2$, where $\hat{k}(x_1,y_1)$ is the longitudinal per-mode loss parameter, as defined
in the main text.  We therefore split the summation over all modes into the components 
\begin{align*}
w_{\|{\rm d}}(x_1,y_1,x_2,y_2,s) &= \sum_{\hat{\omega}=0} \hat{w}_\| (x_1,y_1,x_2,y_2,s)
 , \nonumber \\
w_{\|{\rm c}}(x_1,y_1,x_2,y_2,s) &= \sum_{\hat{\omega}\ne0} \hat{w}_\| (x_1,y_1,x_2,y_2,s)
\end{align*}
and use the causality relation 
$$
   w_{\|{\rm d}}(x_1,y_1,x_2,y_2,s<0)+w_{\|{\rm c}}(x_1,y_1,x_2,y_2,s<0)=0
$$
together with the anti-symmetry of the non-resonant part,
$$
   w_{\|{\rm d}}(x_1,y_1,x_2,y_2,s)=-w_{\|{\rm d}}(x_2,y_2,x_1,y_1,-s) 
$$
proved in (A) below, to eliminate $w_{\|{\rm d}}$, yielding 
$$
w_\|(x_1,y_1,x_2,y_2,s>0) =w_{\|{\rm c}}(x_1,y_1,x_2,y_2,s)+w_{\|{\rm c}}(x_2,y_2,x_1,y_1,-s).
$$
To get the simplified representation for $x_1=x_2$ and $y_1=y_2$, we have to show  that the condition 
$$
\mbox{(B)}\qquad\hat{w}_\|(x_1,y_1,x_1,y_1,s)+\hat{w}_\|(x_1,y_1,x_1,y_1,-s)=-2 \hat{k}(x_1,y_1) \cos(\hat{\omega}s/c)
$$
is fulfilled for eigenmodes with $\hat{\omega}\ne0$.

We will now prove (A) and (B).

\medskip

(A) For \emph{non-oscillating modes} ($\hat{\omega}=0$), the normalized voltage integrals
$\hat{v}(x,y,z)$ are real and the contribution per mode is
$$
\hat{w}_\| (x_1,y_1,x_2,y_2,s)=-2  \int_{-\infty}^\infty
                                                  \hat{v}(x_1,y_1,z+s)
                                                  {\partial \over \partial z} \hat{v}(x_2,y_2,z)
                                   \,\rd z  .
$$
Therefore the required symmetry is fulfilled:
\begin{align*}
\hat{w}_\| (x_1,y_1,x_2,y_2,s)&=-2  \int_{-\infty}^\infty \hat{v}(x_1,y_1,z+s) {\partial \over \partial z} \hat{v}(x_2,y_2,z)  \,\rd z 
\nonumber \\
&=2  \int_{-\infty}^\infty \hat{v}(x_2,y_2,z) {\partial \over \partial z} \hat{v}(x_1,y_1,z+s)  \,\rd z
\nonumber \\
&=2  \int_{-\infty}^\infty \hat{v}(x_2,y_2,z-s) {\partial \over \partial z} \hat{v}(x_1,y_1,z)  \,\rd z
=
-\hat{w}_\| (x_2,y_2,x_1,y_1,-s).
\end{align*}
The physical meaning of this symmetry is that the energy transfer from particle 1 to particle 2
(by $\hat{w}_\| (x_1,y_1,x_2,y_2,s)$) plus the reverse energy transfer
(by $\hat{w}_\| (x_2,y_2,x_1,y_1,-s)$) is zero. This is obvious as no energy is left
to the non-resonant mode after both particles have departed the volume. The voltage $\hat{v}(x,y,z)$ is
zero for $z<0$ before the source $q_1$ entered the cavity and, as the eigensolution is curl-free,
it is zero for $z\ge L$. Therefore $\hat{w}_\| (\cdots\!,s)=0$ for $|s|>L$.
Two particles can  interact only through non-oscillating modes if they are  simultaneously
in the cavity  at any time.

\medskip

(B) The normalized voltage integral for \emph{oscillating modes} ($\omega\ne0$) does not depend on $z$
after $q_1$ has left the cavity:
$$
v(x,y,z>L)=\frac{1}{2 \sqrt{\hat{\cal W}}} \int_{-\infty}^{L} \hat{E}_z(x,y,s) \exp(\ri\, \hat{\omega}s/c) \,\rd s =v(x,y).
$$
Therefore the following integral relation can be derived:
\begin{align*}
\hat{v}^*(x_1,y_1)\hat{v}(x_2,y_2)&=
\int_{-\infty}^{\infty} \frac{\partial}{\partial z} \bigl\{ \hat{v}^*(x_1,y_1,z+s)\hat{v}(x_2,y_2,z) \bigr\} \,\rd z
\nonumber \\
&= \int_{-\infty}^{\infty} \hat{v}(\cdots_2,z) \frac{\partial}{\partial z} \hat{v}^*(\cdots_1,z+s) \,\rd z
   +\int_{-\infty}^{\infty} \hat{v}^*(\cdots_1,z+s) \frac{\partial}{\partial z} \hat{v}(\cdots_2,z) \,\rd z
\nonumber \\
&= \int_{-\infty}^{\infty} \hat{v}(\cdots_2,z-s) \frac{\partial}{\partial z} \hat{v}^*(\cdots_1,z) \,\rd z
   +\int_{-\infty}^{\infty} \hat{v}^*(\cdots_1,z+s) \frac{\partial}{\partial z} \hat{v}(\cdots_2,z) \,\rd z
 . \nonumber
\end{align*}
This relation is needed to prove the symmetry:
\begin{align*}
&\hat{w}(x_1,y_1,x_2,y_2,s)\\
& \quad = -2 \,{\rm Re} \left\{
\exp ( \ri\, \hat{\omega} s/c )  \int_{-\infty}^\infty \hat{v}^*(x_1,y_1,z+s) {\partial \over \partial z} \hat{v}(x_2,y_2,z) \,\rd z 
\right\}
\nonumber \\
&\quad = -2 \,{\rm Re} \left\{
\exp ( \ri\, \hat{\omega} s/c )
\left[\hat{v}^*(x_1,y_1)\hat{v}(x_2,y_2)-
\int_{-\infty}^\infty \hat{v}^*(\cdots_1,z+s) {\partial \over \partial z} \hat{v}(\cdots_2,z) \,\rd z 
\right]
\right\}
\nonumber \\
&\quad  = -2 \,{\rm Re} \bigl\{ \exp ( \ri\, \hat{\omega} s/c )\hat{v}^*(x_1,y_1)\hat{v}(x_2,y_2)\bigr\} -
\hat{w}(x_2,y_2,x_1,y_1,-s)
 . \nonumber
\end{align*}
With $x_1=x_2$ and $y_1=y_2$, we find that 
$$
\hat{w}(x_1,y_1,x_1,y_1,s)+\hat{w}(x_1,y_1,x_1,y_1,-s)=-2 \hat{k}(x_1,y_1) \cos (\hat{\omega}s/c ) ,
$$
where $\hat{k}(x_1,y_1)=|\hat{v}(x_1,y_1)|^2$; in particular, for the origin,
$$
\hat{w}(x_1,y_1,x_1,y_1,0)=-\hat{k}(x_1,y_1) .
$$

    \subsection*{Acknowledgments}
     First and foremost I would like to thank the audience of my lecture
     for their interest. I am also grateful to the CAS team for their organizational efforts
     and patience. Last but not least I am gratefully acknowledging the collaboration
     with Martin Dohlus on a previous CERN school which is the basis for this
     contributions to the proceedings.
%
\bibliography{ntz}

\end{document}